\documentclass[a4paper,11pt]{article}
\pdfoutput=1

\usepackage{jheppub}

\usepackage[T1]{fontenc}

\usepackage{graphicx}
\usepackage{enumitem}
\usepackage{latexsym}
\usepackage{amsfonts}
\usepackage{amssymb}
\usepackage{xcolor}
\usepackage{amsmath}
\usepackage{slashed}
\usepackage{dcolumn}
\usepackage{verbatim}
\usepackage{float}
\usepackage{multirow}
\usepackage{xspace}
\usepackage[normalem]{ulem}
\usepackage[caption=false]{subfig}
\usepackage{bm}
\usepackage{bbm}

\newcommand{\acro}[1]{\textsc{\MakeLowercase{#1}}} 

\renewcommand{\tilde}{\widetilde} 

\newcommand{\beq}{\begin{equation}}
\newcommand{\eeq}{\end{equation}}
\newcommand{\bea}{\begin{eqnarray}}
\newcommand{\eea}{\end{eqnarray}}

\renewcommand{\paragraph}[1]{~\\ \noindent{\bf \emph{#1} --}}

\newcommand{\Fig}[1]{Fig.~\ref{#1}}

\newcommand{\Eq}[1]{Eq.~(\ref{#1})}

\newcommand{\td}[0]{\ensuremath{\text{d}}}

\newcommand{\D}{\mathcal{D}}

\newcommand{\V}{\mathcal{V}}

\newcommand{\bp}{\bar{\phi}}

\newcommand{\T}{\mathcal{T}}

\title{Non-perturbative methods for false vacuum decay}
\author[a,b]{Djuna Croon,}
\author[c,d]{Eleanor Hall,}
\author[c,d,e]{and Hitoshi Murayama}

\affiliation[a]{TRIUMF, 4004 Wesbrook Mall, Vancouver, BC V6T 2A3, Canada
}
\affiliation[b]{Institute for Particle Physics Phenomenology, Department of Physics, Durham University, \\Durham DH1 3LE, U.K.
}
\affiliation[c]{Berkeley Center for Theoretical Physics,\\
University of California, Berkeley, CA 94720, USA}
\affiliation[d]{Theory Group, Lawrence Berkeley National Laboratory, \\
Berkeley, CA 94720, USA}
\affiliation[e]{Kavli Institute for the Physics and Mathematics of the Universe (WPI),\\
University of Tokyo, Kashiwa 277-8583, Japan}

\emailAdd{djuna.l.croon@durham.ac.uk, nellhall@berkeley.edu, hitoshi@berkeley.edu}

\preprint{IPPP/20/95}

\abstract{
    We propose a simple non-perturbative formalism
    for false vacuum decay using functional methods.
    We introduce the quasi-stationary effective action, a bounce action that non-perturbatively incorporates radiative corrections and is robust to strong couplings. The quasi-stationary effective action obeys an exact flow equation in a modified functional renormalization group with a motivated regulator functional. We demonstrate the use of this formalism in a simple toy model and compare our result with that obtained in perturbation theory.
}
\begin{document}
\maketitle
\flushbottom

\section{Introduction}
In the last decade, the arrival of gravitational wave astronomy and the discovery of Higgs metastability have heightened interest in false vacuum (FV) decay.
The stochastic gravitational wave background (SGWB) produced in the aftermath of a primordial first-order phase transition could play a key role in answering open questions in particle physics beyond the standard model (BSM).
With the active development of future gravitational wave detectors, reliably calculating the SGWB for a wide variety of BSM theories has become essential.
These phenomenological studies rely on accurate calculations of the FV decay rate.

The established formalism for FV decay rate calculations, developed by Callan and Coleman \cite{Coleman:1977py,Callan:1977pt}, uses the saddle point approximation to write the imaginary part of the FV energy in terms of bounce solutions of the tree-level action.
Extending this formalism to scenarios in which the tree-level action does not admit bounce solutions has been challenging \cite{Weinberg:1992ds}; and the existing perturbative and non-perturbative methods of incorporating radiative corrections into the bounce formalism are limited to weakly-coupled theories.
With the growing interest in strongly-coupled BSM models, there is a pressing need for new methods for FV decay that are accurate, versatile, and robust to strong couplings.

In this paper, we introduce a new, non-perturbative formalism for FV decay.
We argue that the proper generalization of the saddle-point method defines a \textit{quasi-stationary effective action} (QSEA) which only takes into account fluctuations around the background smaller than a characteristic size.
The QSEA is similar in spirit to the coarse-grained effective action proposed by Langer \cite{Langer:1969bc}, but integrates out small fluctuations in \textit{field configuration} rather than small length scales.

The quasi-stationary effective action may be straightforwardly computed in the language of the non-perturbative functional renormalization group (FRG) \cite{Wetterich:1992yh}.
Within the last decade, the FRG has shown extensive success in tackling non-perturbative problems in QCD; see e.g. \cite{Braun:2008pi,Braun:2009gm,Skokov:2010wb,Herbst:2010rf,Skokov:2010uh,Fister:2011uw,Herbst:2013ail,Mitter:2014wpa,Braun:2014ata,Rennecke:2015eba,Cyrol:2016tym,Fu:2016tey,Cyrol:2017ewj,Cyrol:2017qkl,Fu:2019hdw} and \cite{Dupuis:2020fhh} for a recent review.
We define a slightly modified FRG formulated in terms of the characteristic scale of fluctuations rather than an IR momentum scale.
This modified FRG allows us to write a formally exact flow equation which interpolates between the tree-level action and the quasi-stationary effective action.
We will demonstrate that in the local potential approximation the flow equation may be written as a simple closed-form differential equation that may be straightforwardly solved using ubiquitous numerical tools.

The rest of this paper is organized as follows.
In Sec.~\ref{sec:QSEA_def}, we define the QSEA and discuss how it may solve problems with the perturbative bounce formalism and naive non-perturbative methods.
In Sec.~\ref{sec:FRG}, we show that the QSEA may be implemented in a modified FRG for fluctuations and derive flow equations for the QSEA in the context of the local potential approximation (LPA) with our choice of regulator.
We present numerical results for a real scalar theory at finite temperature.
Finally, we conclude in Sec~\ref{sec:conclusion} by discussing the wide range of new directions, applications, and systematic improvements for the QSEA method.

\section{A quasi-stationary effective action for false vacuum decay}\label{sec:QSEA_def}

In this section, we define the quasi-stationary effective action for FV decay.
In Sec.~\ref{sec:bounce}, we discuss the difficulty of extending the traditional Callan-Coleman ``bounce'' formalism by non-perturbative means.
Whereas naive applications of non-perturbative methods to FV decay are made difficult by the convexity of exact effective actions, careful consideration of the FV decay rate from first-principles yields a non-peturbative, non-convex QSEA that is appropriate for FV decay rate calculations.
In Sec.~\ref{sec:QSEA_def_sub}, we define this QSEA in general terms and use it to write the FV decay rate.

\subsection{The bounce formalism and the convexity of effective actions}\label{sec:bounce}

The bounce formalism uses the stationary phase approximation to write the FV decay rate $\gamma_\text{FV}$ as a sum over saddle points (or stationary points) which correspond to ``bounce'' solutions of the classical Euclidean equations of motion \cite{Coleman:1977py,Callan:1977pt}
\begin{equation}
    \gamma_\text{FV} \simeq A e^{-(S[\phi_b]-S[\phi_F])},
    \label{eq:treelevel}
\end{equation}
where here $S$ is the tree-level action; $\phi_b$ is the bounce solution, which interpolates from the FV at $\tau = \pm\infty$ to a turning point near the true vacuum at $\tau = 0$; $\phi_F$ is the constant false vacuum solution; and the prefactor $A$ includes the fluctuation determinant.
For the purposes of this paper, we will consider a real scalar field theory with renormalizeable Euclidean action
\begin{equation}
    S[\phi] = \int \td^4 x \left[\frac{1}{2} (\partial \phi(x))^2 + V(\phi(x))\right],
\end{equation} 
but our arguments can be generalized to theories with more extensive field content. 

Traditionally, the FV decay rate \eqref{eq:treelevel} was arrived at by directly expanding the imaginary part of the FV energy around the FV, bounce, and multibounce solutions.
However, the path integral for this FV energy is manifestly real, with a non-zero decay rate being a spurious result of perturbation theory; while this can be remedied in an ad-hoc way by meticulously deforming the analytic continuation
\cite{Witten:2010cx}, a first-principles derivation of the decay rate was only put forward in the last decade with the direct method \cite{Andreassen:2016cff}; see \cite{Andreassen:2016cvx} for a comprehensive review.

Without modification, the naive bounce formalism \eqref{eq:treelevel} does not take into account radiative corrections to the action. 
Notably, this implies that it can only be used if bounce solutions exist for the tree-level action, which is not the case for phase transitions generated by radiative corrections.
This signals the breakdown of the stationary phase approximation: when radiative corrections are significant, there exist regions of \textit{bounce-like} field configurations that are not stationary points of the tree-level action but for which the phase varies sufficiently slowly that they nevertheless dominate the integral \cite{Iliopoulos:1974ur,Hindmarsh:1985nc,Fujimoto:1982tc}.
The common way of incorporating radiative corrections has been to apply the bounce formalism to the one-loop (instead of the tree-level) effective action.
Among other issues, this technique is limited to weakly coupled theories as the latter is calculated in perturbation theory.

A related problem is the relation between the bounce formalism and the exact non-perturbative effective action.
While finite-order calculations may give a non-convex effective action, the exact effective action is by definition a convex functional of the fields, and therefore cannot describe the tunneling process of a first order phase transition (see~\cite{Fukuda:1974ey,Weinberg:1987vp,Hindmarsh:1985nc,Fujimoto:1982tc} for relevant discussion).
This may be understood as a consequence of non-local field configurations which are superpositions of the vacuum states \cite{Weinberg:1987vp,Plascencia:2015pga}.
An attractive solution is provided by coarse-graining \cite{Langer:1969bc}: in this method, all modes with momentum above a characteristic scale are integrated out before applying the stationary phase approximation; if the scale is chosen correctly, the resulting action for low momentum modes may be non-convex.
While coarse-graining can be implemented non-perturbatively through dimensional reduction (e.g.~\cite{Ginsparg:1980ef,Appelquist:1981vg,Nadkarni:1982kb,Farakos:1994kx,Braaten:1995cm,Kajantie:1995dw,Croon:2020cgk}) or in the FRG \cite{Alford:1993br,Litim:1996nw,Berges:1996ib,Freire:2000sx,Berges:2000ew,Eichhorn:2020upj}, this procedure requires a scale separation between high and low momentum modes
that breaks down and introduces significant scale dependence in strongly-interacting theories \cite{Berges:1996ib,Croon:2020cgk}.
As a consequence, there is a pressing need for methods robust to strongly interacting theories.  

\begin{figure}[t]
\centering
\includegraphics{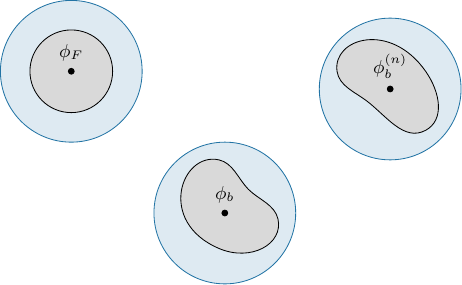}
\caption{Illustration of the integration boundaries in \Eq{eq:Wqsea}. Grey blobs: quasi-stationary regions in field space $U_n$ surrounding quantum solutions of the equations of motion that may contribute to a given path integral, such as the constant false vacuum $\phi_F$, the bounce configuration $\phi_b$, and multibounce solutions $\phi_b^{(n)}$. Blue circles: fluctuations $\varphi$ around the quantum solutions integrated over in the QSEA.}   
\label{fig:qsea}
\vspace{-5mm}
\end{figure}

\subsection{Defining the quasi-stationary effective action}\label{sec:QSEA_def_sub}
Motivated by the picture of radiative corrections as quasi-stationary patches of field space, we propose a new formalism for FV decay based on a non-perturbative generalization of the stationary phase approximation which integrates over local fluctuations in field space.
Under the assumption that the phase varies rapidly everywhere except in certain non-overlapping regions of bounce-like field configurations $U_n$, we may write an arbitrary path integral as
\begin{align}\label{eq:QSEA_int}
    Z \equiv \int \D \phi \,e^{- S[\phi]} \simeq \sum_n \int_{U_n} \D \phi \,e^{- S[\phi]}.
\end{align}
Because the regions outside the dominant patches $U_n$ are negligible by construction, the individual path integrals in \Eq{eq:QSEA_int} are insensitive to the integration boundary, provided it encloses the patch.
This allows the replacement of each individual integration limit $U_n$ with an upper bound $\varphi^2 \lesssim k^{-2} \mathbbm{1}$ on the size of field-space fluctuations $\varphi(x) = \phi(x) - \bp_n(x)$ around the background field $\bp_n = \langle \phi \rangle_n$ within each patch, where here $k$ is some chosen inverse scale.
This condition should be applied in the fluctuation eigenbasis, which for uniform $\bp(x)$ corresponds to the momentum basis, $\varphi(p) \varphi(q) \lesssim k^{-2} \delta(p+q)$.
We may then introduce the \emph{quasi-stationary effective action} (QSEA)
\begin{equation}\label{eq:QSEA}
    \Gamma_k[\bp] \equiv - W_k[J,\bp] + \int J \bar{\phi}\quad  \text{s.t.}\,\,\bp = \frac{\delta W_k}{\delta J}
\end{equation}
in terms of the scale-dependent generator of connected Green's functions
\begin{equation}\label{eq:Wqsea}
    W_k[J,\bp] = \ln \int \D \phi \, e^{- S[\phi] + \int J \phi}\bigg\rvert_{\varphi^2 \lesssim k^{-2} \mathbbm{1}}
    \end{equation}
which only takes into account local fluctuations around $\bp$. With this effective action, the path integral \eqref{eq:QSEA_int} may then be written (up to some path integral normalization $\mathcal{N}$) as
\begin{align}\label{eq:zqsea}
    Z \simeq \mathcal{N} \sum_n \exp[- \Gamma_k[\bp_n]].
\end{align}
As field configurations away from the quasi-stationary points $\bp_n$ yield negligible contributions to the integral by construction, the quasi-stationary effective action becomes large except at those points.
The points $\bp_n$ are thus minima of $\Gamma_k$ and solutions of the \textit{quantum} rather than classical equations of motion $\delta \Gamma_k/\delta \bp = 0$.

The QSEA self-consistently takes into account fluctuations around the quantum solution generated by those same fluctuations, resolving the ``catch 22'' in Ref.~\cite{Croon:2020cgk}.
Although written as a sum around quantum stationary points, the QSEA treats quasi-stationary fluctuations without the need to invoke the stationary phase approximation. 
A representation of the QSEA integration limits in field space is given in \Fig{fig:qsea}.

Because the QSEA only takes into account bounded fluctuations, it is in general non-convex and may admit bounce solutions.
Using the QSEA to evaluate the path integrals in the direct method \cite{Andreassen:2016cff,Andreassen:2016cvx} and correctly accounting for the zero-modes arising from spatial translations of the bounce we arrive at an expression for the FV decay rate
\begin{align}\label{eq:gammaFull}
    \gamma_\text{FV} \simeq 
    \mathcal{J} e^{-(\Gamma_k[\phi_b] - \Gamma_k[\phi_F])},
\end{align}
where within the derivative expansion articulated later in the paper, the Jacobian factor arising from space-time translations of the bounce is given by \cite{Plascencia:2015pga,Andreassen:2017rzq}
\begin{align}
    \mathcal{J} & \simeq \left[\frac{\Gamma_k[\phi_b] - \Gamma_k[\phi_F]}{2 \pi U_k''(
    \phi_F)}\right]^2.
\end{align}
$\Gamma_k$ is analogous to the perturbative bounce action and should converge on the perturbative result in the small coupling limit.
A detailed derivation of the decay rate in the direct method with the QSEA may be found in App.~\ref{sec:decay-derivation}, with further nuances addressed in \cite{direct-forthcoming}.

\section{The functional renormalization group for fluctuations}\label{sec:FRG}
The quasi-stationary effective action may be defined more concretely within the language of the FRG (we include a brief review in App.~\ref{sec:FRG_appendix}).
In Sec.~\ref{sec:FRG_flow}, we introduce a modified FRG that flows in fluctuation size rather than momentum scale, in which the QSEA is naturally realized as the scale-dependent effective action.
In Sec.~\ref{sec:regulators}, we discuss the choice of regulator functionals for this modified FRG and define the particular regulator used in this paper.
In Sec.~\ref{sec:LPA}, we derive the specific flow equations for the QSEA in the LPA, the leading order of the derivative expansion.
Lastly, in Sec.~\ref{sec:results} we present numerical results for the QSEA in the LPA for a real scalar theory at zero temperature.

\subsection{The QSEA flow equation}\label{sec:FRG_flow}
The usual FRG is a powerful tool for non-perturbative problems which introduces a regulator term $R_k(p)$ to the action that makes low-momentum modes below the scale $k$ massive.
An exact flow equation interpolates from the $k\rightarrow \infty$ tree-level action to the full one-particle irreducible (1PI) effective action at $k = 0$ by successively taking into account modes with scale $p\sim k$.\footnote{It was shown in \cite{Litim:2006nn} that this yields a convex effective action.}

Whereas the the characteristic scale $k$ in the traditional FRG acts as a momentum bound, the QSEA scale $k$ is the reciprocal of an upper bound on fluctuation size.
In order to implement the QSEA in the FRG, we introduce a modification to the FRG that flows over fluctuation size rather than momentum scale.
We define a modified regulator functional $R_k[\bp;p]$ which enforces the condition $\varphi^2 \lesssim k^{-2} \mathbbm{1}$ at a scale $k$ by freezing out large fluctuations.
This is in contrast to the usual FRG regulator function, which is $\bp$-independent and freezes out IR modes with momentum $p \lesssim k$.

Within this FRG for fluctuations, we define the QSEA as the scale dependent effective action, which is as usual defined as the modified Legendre transform
\begin{align}
    \Gamma_k[\bp] &= - W_k[J,\bp] + \int J \bp - \Delta S_k[\bp],
\end{align}
with a generator of connected Green's functions $W_k$ that acquires new dependence on $\bp$\footnote{Here $\int_p \equiv \int \td^4 p/(2 \pi)^4$ and $\int_x \equiv \int \td^4 x$}
\begin{align}
    &W_k[J, \bp] = \ln \int \D \phi \exp\left[-S[\phi] + \int J \phi - \frac{1}{2} \int_p R_k[\bp;p] \phi(-p) \phi(p)\right]
    \\
    &\text{and} \,\, 
    \Delta S_k[\bp] = \frac{1}{2} \int_p R_k[\bp;p] \bp(-p) \bp(p).
\end{align}
If the external modified regulator term is absorbed into $W_k[J,\bp]$, its effect is to give mass to the fluctuations away from the mean field.
Although the Legendre transform for $\bp$ is unchanged, the inverse Legendre transform picks up additional pieces due to the $\bp$-dependence of the regulator functional and must therefore be defined implicitly
\begin{align}\label{eq:JQSEA}
    \bp &= \frac{\delta W_k[J,\bp]}{\delta J} \equiv W_k^{(1,0)}[J,\bp] ,
    \\
    J &= \Gamma_k^{(1)}[\bp] + W_k^{(0,1)}[J,\bp] + \Delta S_k^{(1)}[\bp],
\end{align}
where the superscripts indicate partial functional derivatives.
The scale-dependent effective action obeys the formally exact flow equation,
\begin{align}\label{eq:floweq}
    \partial_k \Gamma_k[\bp]
    = \frac{1}{2} \int_p (\partial_k R_k) G_k[\bp;-p,p],
\end{align}
in terms of the exact propagator for the theory at the scale $k$,
\begin{align}
    G_k[\bp] = W_k^{(2,0)}[J,\bp] = \frac{\delta \bp}{\delta J}\ .
\end{align}
%
The full expression for the propagator,\footnote{The terms $W_k^{(1,1)}$, which arise due to the new functional dependence of $W_k$ on $\bp$ in the QSEA, were omitted from Eq.~\eqref{eq:propagator} in the version of this paper which originally appeared on the arXiv. As discussed in App.~\ref{sec:propagator_new_terms}, they ultimately do not affect our results.}
\begin{align} \label{eq:propagator}
    G_k[\bp;p,q] =
    \int_{p',q'} (\mathbbm{1} - W_k^{(1,1)})_{p,p'}^T \left[\Gamma_k^{(2)} + W_k^{(0,2)}+ \Delta S_k^{(2)}\right]_{p',q'}^{-1} (\mathbbm{1} - W_k^{(1,1)})_{q',q},
\end{align}
is derived in App.~\ref{sec:propagator_new_terms},
where here $[\dots]^{-1}$ denotes a functional inverse and the simplified notation $(\mathbbm{1} - W_k^{(1,1)})_{p,q} = \delta(p-q) - \frac{\delta^2 W_k}{\delta \bp(p) \delta J(q)}$ with $\delta /\delta \bp$ and $\delta /\delta J$ here indicating \textit{partial} functional derivatives.
Although this expression is not in general closed-form, we will show that it admits closed-form solutions for $G_k$ for certain well-motivated choices of regulator and within the derivative expansion, which can be substituted directly into the flow equation, \Eq{eq:floweq}.

The exact flow equation flows from the initial condition $\Gamma_\Lambda = S$ at a UV scale $k = \Lambda$ to the QSEA at some scale $k_\text{min}$.
A natural question is how to choose the correct scale $k_\text{min}$ at which to stop the flow, which corresponds to the inverse of the maximum scale of fluctuations incorporated into the QSEA.
In contrast with earlier proposals \cite{Berges:1996ib,Berges:2000ew}, the QSEA with an appropriately defined regulator functional 
(as in the following section) converges to an action which is manifestly non-convex in the limit $k_\text{min} \to 0$.
The QSEA stops being sensitive to the scale $k$ when all bounce patches are fully enclosed, such that any choice of $k_\text{min}$ below the scale of convergence yields consistent results.
Therefore we are free to relax the scale all the way to $k_\text{min}=0$ in the following, which amounts to taking into account all \textit{local} fluctuations (corresponding to $S_k^{(2)}[\bp] \geq 0$) around the mean field.

\subsection{Regulator functionals for fluctuation size}\label{sec:regulators}
As motivated above, the QSEA can be incorporated into FRG by an appropriate definition of the regulator that constrains the size of fluctuations $\varphi^2 \lesssim k^{-2}\mathbbm{1}$. 
The leading inverse scale of quantum fluctuations around a constant mean-field $\bp$ is given by the fluctuation operator $-\partial^2 + V_k''$ for the scale dependent theory $V_k'' = V'' + R_k$. The eigenbasis for the fluctuation operator is the momentum basis $\varphi(p)$, with eigenvalues $p^2 + V_k''$. Therefore, for uniform $\bp(x) = \bp$ as encountered in the context of derivative expansion later in the paper we will focus on a class of regulator functionals $R_k$ which satisfy the criteria
\begin{align}
    p^2 + V'' \ll k^2 &\implies R_k[\bp;p] \sim k^2 ,
    \\
    p^2 + V'' \gg k^2 &\implies R_k[\bp;p] \sim 0 ,
\end{align}
such that $p^2 + V_k'' \gtrsim k^2$ everywhere.
For the purpose of the local potential approximation presented in this paper, we will choose a regulator
\begin{align}
    R_k[\bp;p] &= \left(k^2 - \frac{1}{\V} \int_x \left[p^2 + V''(\bp(x))\right]\right) \cdot\Theta \left(k^2 - \frac{1}{\V} \int_x \left[p^2 + V''(\bp(x))\right]\right),
    \label{eq:regulator}
\end{align}
which for uniform $\bp(x)$ may be seen as a minimal choice: it sets $p^2 + V_k'' = k^2$ for $p^2 + V'' < k^2$ and has no effect when $p^2 + V'' \geq k^2$. Here and below, $\Theta(x)$ is the Heaviside (step) function, and $\V$ spacetime volume.
Although this regulator was chosen to correctly regulate the scale of fluctuations, we will show that within the derivative expansion it also gives closed-form solutions to \Eq{eq:propagator}, dramatically simplifying the task of solving the flow equation.
Outside of the local potential approximation, the choice of regulator becomes more complex, as the fluctuation eigenbasis is no longer the momentum basis; this is the subject of ongoing work.

It is worth noting that one may use any regulator function $R_k$ which consistently takes into account fluctuations $\sim k^{-2}$ at the scale $k$, because the QSEA is insensitive to the integration boundary by construction.
Other well-motivated choices might substitute $\Gamma_{k_\text{min}}^{(2)}$ in place of the average fluctuation operator in \Eq{eq:regulator}, or use the averaged field rather than the averaged fluctuation operator.
These choices become more important outside of the derivative expansion, and will be the subject of future works.
One might even be tempted to make a {\it field-independent} choice analogous to \eqref{eq:regulator} evaluated at a fixed background field $\bp_0$, as has been explored in Ref.~\cite{Litim:2002cf}; one specific example would be $R_k = (k^2 - p^2 - \min_{\phi} V'')\Theta(k^2 -p^2 - \min_{\phi} V'')$.
This choice of regulator is equivalent to coarse-graining with the scale $k_\text{min}^2 = \min V''(\phi)$.
Although it does enforce $p^2 + V'' \geq k^2$ everywhere, this approximation \textit{smears} the size of field-space fluctuations, so that the fluctuation size is regulated inconsistently across different field values.
Smearing the fluctuation size results in scale-dependence of the coarse-grained effective action: it is not in general possible to take into account all local fluctuations without also incorporating non-local fluctuations.
As has been observed \cite{Berges:1996ib}, this effect is expected to worsen for strongly-coupled theories.

Different choices of regulator might lead to slight variations in result under the various approximation schemes used to solve the flow equation, as well as different convergence behavior.
In particular, one might imagine using convergence criteria to define an optimized regulator as has been done in the conventional FRG \cite{Litim:2000ci,Litim:2001up}.
Such an analysis is beyond the scope of the present paper and will be deferred to future work.

\subsection{The local potential approximation}\label{sec:LPA}
Although the flow equation is formally exact, in practice approximation schemes are required to solve it. Reasonable approximation schemes are fully robust to strongly-coupled systems and preserve the one-loop structure of the flow equation.
For the purposes of this paper, we will work within the leading order of the derivative expansion (DE), known as the local potential approximation (LPA).\footnote{While the regulator used in this paper is only valid in the context of the DE, the QSEA is not in principle limited to the DE or the LPA; extending beyond the LPA is the subject of ongoing work.}
In the LPA, one takes an ansatz for the effective action
\begin{align}\label{eq:LPA_ansatz}
    \Gamma_k[\bp] = \int \td^4 x \left[\frac{1}{2} (\partial \bp)^2 + U_k(\bp(x))\right]
\end{align}
in which all the scale-dependence is contained in a local effective potential $U_k(\bp(x))$.
Evaluating both sides of the flow equation for $\Gamma_k$ at constant $\bp$ then yields a flow equation for the quasi-stationary effective potential $U_k$. 

\begin{figure*}[t]
\centering
\subfloat[\label{fig:weak}]{
\includegraphics{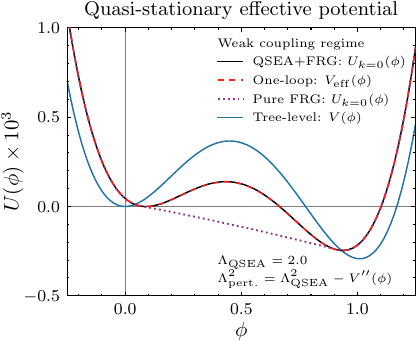}
}
\hfill
\subfloat[\label{fig:strong}]{
\includegraphics{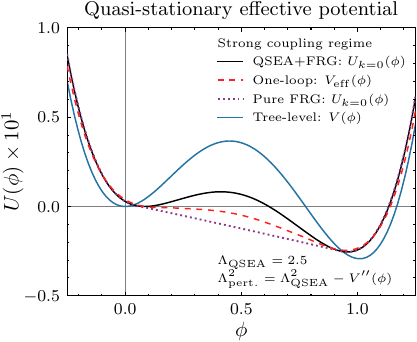}
}
\caption{The QSEA in action. The effective potential calculated using the QSEA flow equation in the LPA \Eq{eq:flowLPA} (black), in 4D perturbation theory (red dashed), and in the unmodified FRG (purple dotted) for the tree-level potential \Eq{eq:toypotential} (blue). {\bf(a)}~Weakly-coupled theory with $m^2 = 0.014$, $\alpha =-0.09$, $\lambda =0.185$, with UV-scale $\Lambda_{\rm QSEA} = 2$ for units $v = 1$. {\bf(b)}~Strongly-coupled theory with $m^2 = 1.4$, $\alpha = -9$, $\lambda = 18.5$, with UV-scale $\Lambda_{\rm QSEA} = 2.5$. As expected, the QSEA exhibits minimal scale dependence: by $k = 0.1$ the QSEA has converged to $|U_{k=0.1}(\phi) - U_{k=0}(\phi)| \leq 3.6 \cdot 10^{-7}$ for the parameters in (a) and $1.9 \cdot 10^{-5}$ for the parameters in (b). 
}
\label{fig:toymodel}
\end{figure*}

With the regulator specified in \Eq{eq:regulator}, the constant-field propagator dramatically simplifies, as we will demonstrate.
As discussed in App.~\ref{sec:propagator_new_terms}, the $W_k^{(1,1)}$ terms in the propagator only contribute $\propto \delta(p) \delta(q)$.
Then, carefully expanding the propagator for uniform $\bp$ and performing the inversion as in App.~\ref{sec:propagator_LPA}, the result may be summarized as
\begin{align}
    G_k[\bp;p,q] &= G_k(\bp) \delta(p+q) + \{\text{terms } \propto \delta(p) \delta(q) \} 
    \label{eq:Gkpq}
\end{align}
for $G_k(\bp)$ that is a function of the constant field value 
\begin{equation}\label{eq:Gphi}
\begin{split}
    G_k(\bp) = \frac{1}{p^2 + U_k'' + R_k + \frac{V''''}{2} \int_{p'} \hspace{-0.6em} \frac{G_k[\bp;-p',p']}{\V} \Theta(\tilde{k}^2 - p'^2)}
\end{split}
\end{equation}
rather than a functional of the entire field configuration and terms $\delta(p)\delta(q)$ that may be excluded by choosing the regulator to vanish at this point. 
Here for convenience we have introduced $\tilde{k}^2(\phi) \equiv k^2 - V''(\phi)$.
In both the flow equation~\eqref{eq:floweq} and the above~\eqref{eq:Gphi}, the propagator is always associated with a Heaviside function $\Theta(\tilde{k}^2 - p^2)$. In this context, resolving the $\Theta$-function in $R_k $ yields a $G(\bp)$ that is $p$-independent over the integration range. We may therefore evaluate the integral to find
\begin{align}
    G_k(\bp) = \frac{1}{\tilde{k}^2 + U_k'' + \frac{\tilde{k}^4(\bp)}{64 \pi^2} V'''' G_k(\bp)}.
    \label{eq:GphiInt}
\end{align}
Eq.~\eqref{eq:GphiInt} has the closed-form solution
\begin{equation}
    G_k(\bp) = \frac{1}{\tilde{k}^2(\phi) + U_k''(\phi)} \frac{2}{1 + \sqrt{1 + \frac{V''''(\phi) \tilde{k}^4(\phi)}{16 \pi^2 (\tilde{k}^2 + U_k'' )^2}}}\ .
    \label{eq:closed_form}
\end{equation}
Lastly, we may write the flow equation in the LPA
\begin{equation} \label{eq:flowLPA} \begin{split}
    \partial_k U_k(\bp) =& \int_p k \Theta(\tilde{k}^2(\bp) - p^2) \frac{G_k[\bp;-p,p]}{\V}
    \\
    =& \,\frac{k \tilde{k}^4(\bp) \Theta(\tilde{k}^2)}{32 \pi^2 (\tilde{k}^2 + U_k'')} \frac{2}{1 + \sqrt{1 + \frac{V''''(\phi) \tilde{k}^4(\phi)}{16 \pi^2 (\tilde{k}^2 + U_k'')^2}}} \ .
    \end{split}
\end{equation}
Further discussion of the inversion of the propagator and the derivation of the flow equation may be found in App.~\ref{sec:propagator_LPA}.
\subsection{Demonstration in a simple model}\label{sec:results}
We will consider a model with one real scalar field with potential
\begin{align} \label{eq:toypotential}
    V(\phi) = \frac{m^2 }{2} \phi^2 + \frac{\alpha}{3!} \phi^3 + \frac{\lambda}{4!}  \phi^4.
\end{align}
For appropriate values of $m^2,\alpha,\lambda$ with a negative cubic term this theory has two minima, such that FV decay may occur from the local to the global minimum.

\Fig{fig:toymodel} shows the effective potential for two sets of parameters. 
In \Fig{fig:weak}, the QSEA is evaluated for parameters well into the perturbative regime.
%
We compare the result to the one-loop effective potential and the unmodified (``pure'') FRG with the bare parameters $m^2,\alpha,\lambda$ defined at the UV momentum cutoff $p_\text{max}^2(\phi) = \Lambda^2 - V''(\phi)$,
arising from the definition of the QSEA-FRG in terms of fluctuation size.
An alternative scheme is to transform the flow equation back to a momentum scale and perform the usual normalization.
We observe that, as expected, the QSEA agrees nearly identically with the one-loop effective potential.

In \Fig{fig:strong}, the tree-level parameters are in the strong coupling regime. 
As expected, the non-perturbative QSEA and the perturbative effective potential disagree significantly.
In fact, the perturbative result gives \textit{qualitatively} different behavior than the QSEA: whereas the QSEA predicts a first-order phase transition, the perturbative result does not.
This difference underscores the importance of non-perturbative methods which are robust in the strong-coupling regime.
\section{Conclusion}\label{sec:conclusion}
In this work we have proposed a new non-perturbative technique to study false vacuum decay through the quasi-stationary effective action defined by \Eq{eq:QSEA}.
The QSEA takes into account local fluctuations in field-space which dominate the path integral, resulting in an expression for the false vacuum decay rate without fluctuation determinant, \Eq{eq:gammaFull}.
The formalism has a natural implementation within the framework of functional renormalization: the locality of fluctuations can be enforced through a false vacuum regulator in a subtly modified functional renormalization group.
This technique allows for very simple numerical application using a variety of integration techniques. 
The success of QSEA is demonstrated for a simple model defined by \Eq{eq:toypotential}, through a direct comparison with the result obtained in 1-loop perturbation theory in Fig.~\ref{fig:toymodel}.

A recent alternative proposal for the calculation of the FV effective action is based on the two-particle irreducible (2PI) effective action \cite{Garbrecht:2015cla,Garbrecht:2015yza}.
Similar methods may also be applied within the context of the FRG, defining an alternative flow equation in which the regulator plays a very different role \cite{Alexander:2019cgw,Alexander:2019quf}.
FV decay in the latter context has not been studied yet, but it will be interesting to compare the results with those found using the techniques proposed here.
Indeed, the QSEA introduced in this paper may be naturally understood in the language of the 2PI \cite{2PI-forthcoming}, yielding FRG flow equations including additional mass dressing terms; this 2PI-QSEA may allow for more robust handling of higher orders of the derivative expansion which become urgent at finite temperature.

Our results encourage further study of the QSEA in a range of directions. 
The derivation for a single real scalar field can be expanded to include coupled degrees of freedom of different spin; the inclusion of finite temperature into the flow equation is straightforward. 
Future effective field theory studies of the electroweak phase transition along the lines of \cite{Croon:2020cgk} may determine whether the resulting theoretical uncertainty is reduced. 
Moreover, the QSEA could be a useful tool in studies of the zero-temperature stability of the electroweak vacuum \cite{Gies:2017ajd}.

Importantly, the techniques introduced here can be applied to strongly interacting theories. A key strength of functional renormalization is its ability to non-perturbatively treat large couplings and interpolate between IR and UV pictures of theories with composite scalars. 
The Polyakov-loop enhanced quark meson model (PQM) \cite{Fukushima:2003fw} has a natural implementation in the FRG \cite{Braun:2009gm,Skokov:2010wb,Herbst:2010rf,Skokov:2010uh,Schaefer:1999em,Bohr:2000gp,Herbst:2013ail}, and the techniques derived in this work may be extended to this model.
Future work along this direction will be aimed at studying (thermal) phase transitions occurring in strongly coupled models of dark matter (e.g.~\cite{Hochberg:2014kqa,Tsumura:2017knk,Aoki:2017aws,Bai:2018dxf,Helmboldt:2019pan,Hall:2019rld}), models with heavy axions~\cite{Croon:2019iuh} and QCD baryogenesis~\cite{Ipek:2018lhm,Croon:2019ugf}.
In these models, QSEA is the first step towards reliable predictions of the stochastic gravitational wave background from the early universe.

\acknowledgments
The authors are grateful to Rachel Houtz and Peter Millington for useful conversations and Oli Gould and Mark Hindmarsh for feedback on the manuscript.
T\acro{RIUMF} receives federal funding via a contribution agreement with the National Research Council Canada.
The work of EH was supported by the NSF GRFP.
DC and EH thank the Aspen Center for Physics, supported by NSF grant PHY-1607611, for (virtual) hospitality during the completion of this paper, and all participants of the workshop ``A Rainbow of Dark Sectors" for important work and stimulating conversations.
The work of HM was supported in part by the DOE under grant DE-AC02-05CH11231, in part by the NSF grant
PHY-1915314, PHY-2210390, by the JSPS Grant-in-Aid for
Scientific Research JP20K03942, MEXT Grant-in-Aid for Transformative Research Areas (A)
JP20H05850, JP20A203, by WPI, MEXT, Japan, Beyond AI Institute at the University of Tokyo, and Hamamatsu Photonics, K.K.

\appendix

\section{Functional renormalization}\label{sec:FRG_appendix}
Here we review a few important aspects of the unmodified functional renormalization group (FRG), but we encourage the unfamiliar reader to learn the basics of the methodology elsewhere, for example through \cite{Wetterich:1992yh,Dupuis:2020fhh}. The FRG works through the introduction of a scale-dependent IR cutoff term to the action
\begin{equation}
    \Delta S_k[\bp] = \frac{1}{2} \int_p R_k(p) \bp(-p) \bp(p).
\end{equation}
Here $R_k (p)$ is the \emph{regulator function}, and must be chosen such that low-momentum modes $p < k$ below the FRG scale $k$ are made massive, while high-momentum modes $p > k$ above the FRG scale are not. A typical choice of regulator is the theta-function regulator $R_k \propto \Theta(k^2 - p^2)$ \cite{Litim:2000ci,Litim:2001up}.
With this term, we can define a scale-dependent effective action as the modified Legendre transform
\begin{align}
    \Gamma_k[\bp] &= -W_k[J] + \int J \bp - \Delta S_k[\bp] ,
    \\
    W_k[J] &= \ln \int \D \phi \, e^{-S[\phi] + \int J \phi - \Delta S_k[\phi]}.
 \end{align}
The scale dependent effective action interpolates between the microscopic action at $k \to \infty$, where all modes are frozen out, and the exact 1PI effective action at $k = 0$, where all modes are taken into account and $\Delta S_k = 0$. 
The effective action at different scales related by a formally exact flow equation \cite{Wetterich:1992yh}
\begin{align}\label{eq:floweqsup}
    \partial_k \Gamma_k[\bp]
    = \frac{1}{2} \int_p (\partial_k R_k(p)) G_k[\bp;-p,p],
\end{align}
where 
\begin{equation}
    G_k[\bp;p,q] = \left[\Gamma_k^{(2)}[\bp] + R_k \delta(p+q)\right]^{-1}
\end{equation} 
is the exact propagator.
This equation is also called the Wetterich equation \cite{Wetterich:1992yh}.

To solve the flow equation \Eq{eq:floweqsup} in practice, approximations such as the derivative expansion and vertex expansion (distinct from a weak coupling expansion) must be used.
The derivative expansion employs a local ansatz for the effective action expanded to orders in the derivative; the leading-order LPA employed in this paper takes the ansatz \eqref{eq:LPA_ansatz} including a single derivative term and a local potential.
In the vertex expansion, the coupled flow equations for the various 1PI vertices are truncated at some order $n$ s.t. $\partial_k \Gamma^{(n)}[\bp] = 0$.
The derivative expansion
is valid for small momenta $p\leq \rm{max}(k,\xi^{-1}) $, where $\xi $ is the correlation length; within the context of the FRG the derivative expansion is particularly well motivated because the regulator function $R_k(p)$ enforces $p \lesssim k$ within the flow equation.
With optimized regulator functions \cite{Litim:2000ci,Litim:2001up} chosen to maximize the convergence of these approximation schemes, the derivative expansion has been found to be highly effective for many applications even at leading order; see e.g. \cite{Braun:2009gm,Skokov:2010wb,Herbst:2010rf,Skokov:2010uh,Schaefer:1999em,Bohr:2000gp,Herbst:2013ail} for applications within QCD. 

\section{The decay rate in the direct method}\label{sec:decay-derivation}
In the direct method~\cite{Andreassen:2016cff}, the decay rate can be written as
\begin{align}
    \label{eq:rate}
    \gamma_\text{FV} = \frac{1}{\V_3}\lim_{T \rightarrow \infty} \left\lvert 2 \, \text{Im} \, \left(\frac{\int \D \phi \, e^{- S_E[\phi]} \delta(\tau_\Sigma[\phi])}{\int \D \phi \, e^{- S_E[\phi]}}\right)_{\substack{\T >0 \\ \T =i T}}\right\rvert ,
\end{align}
where $\tau_\Sigma$ is a functional returning the imaginary time at which the field configuration crosses the turning-point surface $\Sigma$.
In the above, the crucial analytic continuation is contained within the expression $\substack{\T > 0 \\ \T = i T}$. As per the discussion below Eq.~(4.15) in Ref.~\cite{Andreassen:2016cvx}, this notation is used with a very specific meaning. First, the path integrals in both the numerator and denominator are evaluated for real $\T >0$ (i.e. in Euclidean space). Following the evaluation of the path integrals, the expression inside the parentheses is then analytically continued to $\T = i T$ (i.e. back to Minkowski space). Lastly, the imaginary part is taken. 
The dependence on $\T$ may be made more explicit by introducing a phase factor $r$ as per Eq.~(4.15) in Ref.~\cite{Andreassen:2016cvx},
\begin{align}
    \label{eq:rate_r}
    \gamma_\text{FV} = \frac{1}{\V_3}\lim_{T \rightarrow \infty} \left\lvert 2 \, \text{Im} \, \left(\frac{\int \D \phi \, e^{- r S[\phi]} \delta(r t_\Sigma[\phi])}{\int \D \phi \, e^{- r S[\phi]}}\right)_{\substack{r > 0 \\ r =i}}\right\rvert .
\end{align}
Here as before we evaluate the path integrals in Euclidean space for real $r>0$, analytically continue to $r = i$, and then take the imaginary part.
This procedure is introduced on pages 34 and 35 of Ref.~\cite{Andreassen:2016cvx}, and is derived from first principles.
Typically, this expression is evaluated using the saddle-point approximation for both the numerator and the denominator; to go beyond perturbation theory, we will instead use the non-perturbative generalization of the saddle-point approximation introduced in this paper.
The numerator is dominated by the stationary patch $U_b$ surrounding the bounce solution, whereas the denominator is dominated by the stationary patch $U_F$ surrounding the false vacuum.

In dealing with these regions, we must first treat the zero modes $\partial_\mu \bp_b$ corresponding to translations of the bounce. 
We will follow a prescription which largely follows \cite{Andreassen:2017rzq}. We can write a general field configuration over some basis $\phi_i(x)$ as
\begin{align}
    \phi = \bp_b(x) + \xi_i \phi_i(x) .
\end{align}
In the typical perturbative case, this basis is chosen to be the eigenbasis of $S^{(2)}$; however because we are working in terms of quantum solutions, we will work in the eigenbasis of $\Gamma^{(2)}$, with eigenvalues $\lambda_i$. We will choose the fields to be normalized according to the convention
\begin{align}
    \langle \phi_i \vert \phi_j \rangle = \int_x \phi_i \phi_j = 2 \pi \delta_{i j} .
\end{align}
The space-time translation modes $\partial_\mu \bp_b$ can be shown to have zero eigenvalue. These modes may be accounted for via a collective coordinate, up to a Jacobian factor which we will now determine. The inner product of the translation modes, without normalization is
\begin{align}
    \langle \partial_\mu \bp_b \vert \partial_\nu \bp_b \rangle = \int_x (\partial_\mu \bp_b) (\partial_\nu \bp_b) = \delta_{\mu \nu} \int_x (\partial_\mu \bp_b)^2 \quad \text{(no sum)} .
\end{align}
The properly normalized zero modes are thus $\sqrt{2 \pi/ \int_x (\partial_\mu \bp_b)^2}\, \partial_\mu \bp_b$. Therefore, we can separate out the zero modes in the definition of the general field configuration as
\begin{align}
    \label{eq:zeromodes}
    \phi = \bp_b(x) + \sum_\mu \xi_\mu \sqrt{\frac{2 \pi}{\int_x (\partial_\mu \bp_b)^2}}\, \partial_\mu \bp_b +  \sum_i \xi_i \phi_i(x) .
\end{align}
An equivalent parameterization is in terms of a collective coordinate $x_0$ to account for the space-time translations of the bounce corresponding to $\lambda_0=0$. Within this alternate formulation, we should write
\begin{equation}
    \phi = \bp_b(x + x_0) + \zeta_i \phi_i(x + x_0) .
\end{equation}
Expanding for small $x_0$, we observe
\begin{align}
    \label{eq:collective}
    \phi = \bp_b(x) + x_0^\mu \partial_\mu \bp_b(x) + \zeta_i \phi_i(x) + \zeta_i x_0^\mu \partial_\mu \phi_i(x) .
\end{align}
Comparing Eqs.~\eqref{eq:zeromodes} and~\eqref{eq:collective}, we do indeed recover the same modes $\partial_\mu \bp_b$ as promised, up to a Jacobian factor.
Note that there is an important distinction between the sum over $\mu$ in the above expression and \Eq{eq:zeromodes}: in the above, the sum should include the metric $x_0^\mu \partial_\mu \bp_b = g^{\mu \nu} x_\mu \partial_\mu \bp_b$, whereas in \Eq{eq:zeromodes} it should not.
This makes no difference for a Euclidean collective coordinate; however, if one wished to use a general time coordinate (e.g. Minkowski time) there would be additional phases.
As a consequence, when performing the analytic continuation in \eqref{eq:rate_r}, the collective coordinate integral and Jacobian we have defined here and below do not rotate with $r$. 
If one were to use a collective coordinate that did rotate with $r$, any phase would cancel between the integral measure and the Jacobian.
For small $\zeta$ and a Euclidean collective coordinate, the Jacobian between $\zeta_i$ and $\xi_i$ is one and the Jacobian between $x_0$ and $\xi_\mu$ is
\begin{align}
    \label{eq:jacobian}
    \mathcal{J}[\bp_b] = \prod_\mu \sqrt{\frac{\int_x (\partial_\mu \bp_b)^2}{2 \pi}} \quad \text{(no sum)}.
\end{align}
Next, we will evaluate the Jacobian term in Euclidean space with real $\T > 0$. We know that, because of the spherical symmetry of the bounce, before analytic continuation from real $\tau$ to imaginary $\tau$ it's true that
\begin{align}
    \int_x (\partial_0 \bp_b)^2 = \int_x (\partial_i \bp_b)^2 \,\, \text{(no sum)} = \Gamma[\bp_b] - \Gamma[\bp_F] .
\end{align}
This arises from invariance under infinitesimal coordinate dilations. Because action is invariant under $\bp(x) \rightarrow \bp(e^{-a} x)$ for $a \rightarrow 0$, we know that the variation of the action under such transformations is zero. We can then use the fact that
\begin{align}
    \lim_{a \rightarrow 0} \frac{\partial \partial_\mu \bp}{\partial a} = - \partial_\mu \bp - x_\nu \partial_\mu \partial_\nu \bp
\end{align}
to write the variation of the action as
\begin{align}
    \delta \Gamma = 0 &= \int_x \left[-(\partial_\mu \bp)^2 - (\partial_\mu \bp) (x_\nu \partial_\mu \partial_\nu \bp) - (x_\nu \partial_\nu \bp) U'(\bp)\right]
    \\
    &= \int_x \left[-(\partial_\mu \bp)^2 - \frac{1}{2} x_\nu \partial_\nu (\partial_\mu \bp)^2 - x_\nu \partial_\nu U(\bp)\right] .
\end{align}
If the potential is normalized so that $U(\bp_F) = 0$ (i.e. replacing $\Gamma$ with $\Gamma - \Gamma_F$) and noting that that $\partial_\mu \bp$ vanishes at the integral boundary, we can use integration by parts on the last two terms
\begin{align}
    \delta \Gamma
    &= \int_x \left[-(\partial_\mu \bp)^2 + 2 (\partial_\mu \bp)^2  + 4 U(\bp)\right]
    \\
    &= \int_x \left[(\partial \bp)^2  + 4 U(\bp)\right] = 0
\end{align}
with the factor of $4$ coming from the sum. Therefore, returning to our definition of the effective action,
\begin{align}
    \Gamma[\bp_b] - \Gamma[\bp_F] &= \int_x \left[\frac{1}{2} (\partial \bp_b)^2 + U(\bp_b) - U(\bp_F) \right]
    \\
    &= \int_x \left[\frac{1}{2} (\partial \bp_b)^2 - \frac{1}{4} (\partial \bp_b)^2 \right]
    \\
    &= \frac{1}{4} \int_x (\partial \bp_b)^2 .
\end{align}
Lastly, the spherical symmetry of the bounce means that in Euclidean coordinates
\begin{align}
    \frac{1}{4} \int_x (\partial \bp_b)^2 = \int_x (\partial_0 \bp)^2 = \int_x (\partial_i \bp)^2 \text{ (no sum)} = \Gamma[\bp_b] - \Gamma[\bp_F].
\end{align}
This allows us to write the Jacobian for a Euclidean collective coordinate as
\begin{align}
    \mathcal{J}[\bp_b] = \left(\frac{\Gamma[\bp_b] - \Gamma[\bp_F]}{4 \pi^2}\right)^2
\end{align}
which is real. We should also perform a similar treatment for translations of the shot.

With this Jacobian factor, we can extract the collective coordinates from the path integral
\begin{equation}
    \begin{split}
        \int \D \phi \, e^{- r S[\phi]} \delta(r t_\Sigma [\phi]) \simeq \int \td^4 x_0 \bigg[&\int_{U_b} \D' \phi \, \mathcal{J}[\phi] e^{- r S [\phi]} \delta(r t_\Sigma[\phi]) 
        \\
        + \, & \int_{U_s} \D' \phi \, \mathcal{J}[\phi] e^{- r S[\phi]} \delta(r t_\Sigma[\phi])\bigg],
    \end{split}
\end{equation}
where the integrals are restricted to the regions $U_b$ and $U_s$ surrounding the quantum bounce and shot solutions, respectively, and where $ \Sigma$ denotes the turning-point surface.
The measure $\D'\phi$ indicates that the zero modes have been extracted from the path integral.
While the restriction that the integrals run over $U_b$ and $U_s$ mean that these translational modes don't show up in the integration, the integrals $\int_U \D \phi$ and $\int_U \D' \phi$ are dimensionally different as the latter has the second through fifth eigenmodes omitted.
In order to account for this difference, we will extract the corresponding eigenmodes from the denominator.\footnote{Equivalently, we could also shift the eigenmodes in the numerator over four places so that in the numerator $\{\lambda_1, \lambda_6,\lambda_7,\dots\} \rightarrow \{\lambda_1,\lambda_2,\lambda_3\}$.}
The collective coordinate integral over $t_0$ then simply resolves the delta-function, yielding a $1/r$ factor and fixing $t_0$ to some value $t^*$ if $\phi$ indeed crosses $\Sigma$ and otherwise simply yielding $0$; we will summarize this with a term $\Theta_\Sigma[\phi]$ that yields $1$ if $\phi$ hits $\Sigma$ and $0$ otherwise,
\begin{align}
    \left(\int d t_0 \, \delta(r t_\Sigma[\phi])\right)_{r > 0} = \frac{1}{r} \Theta_\Sigma[\phi] .
\end{align}
Note that here, as $r > 0$, $t_0$ is a Euclidean coordinate.
The spatial integral over $\vec{x}_0$ simply yields a volume factor, and so
\begin{align}
    \int \D \phi \, e^{- r S[\phi]} \delta(r t_\Sigma [\phi]) &\simeq
    \frac{\V_3}{r} \left[\int_{U_b} \D' \phi \, \mathcal{J}[\phi] e^{- r S[\phi]} \Theta_\Sigma[\phi] + \int_{U_s} \D' \phi \, \mathcal{J}[\phi] e^{- r S[\phi]} \Theta_\Sigma[\phi]\right] .
\end{align}
Here and below, $\V_3$ is the volume of three-dimensional space. Because we are evaluating over only local regions around the bounce and shot, we can pull the Jacobian factors out of the integral as well evaluate the $ \Theta_\Sigma$ term, leading to a factor of $1/2$ for the bounce term due to the fact that only half of the fluctuations around the bounce solution satisfy $\Theta_\Sigma$,
\begin{align}
     \int \D \phi \, e^{- r S[\phi]} \delta(\tau_\Sigma [\phi]) &\simeq \frac{\V_3}{r} \left[\frac{1}{2} \mathcal{J}[\bp_b] \int_{U_b} \D' \phi\, e^{- r S[\phi]} + \mathcal{J}[\bp_s] \int_{U_s} \D' \phi\, e^{- r S[\phi]} \right] .
\end{align}
Finally, we may  introduce the quasi-stationary effective action for the integrals around the bounce and shot regions to get the numerator
\begin{align}
    \int \D \phi \, e^{- r S[\phi]} \delta(r t_\Sigma [\phi]) &\simeq \mathcal{N} \frac{\V_3}{r} \left[\frac{1}{2} \mathcal{J}[\bp_b] e^{-\Gamma_\text{QSEA}[\bp_b]} + \mathcal{J}[\bp_s] e^{-\Gamma_\text{QSEA}[\bp_s]}\right] .
\end{align}
For the denominator, we don't have to deal with this problem of collective coordinates; however, in order for the integration measures to align we should extract eigenmodes 2-5, which have eigenvalue $\mathcal{M} = \sqrt{U''(\phi_F)}$. Therefore, we will get
\begin{align}
    \int \D \phi \, e^{-r S[\phi]} = \mathcal{M}^4 \int \D' \phi \, e^{-r S[\phi]} \simeq \mathcal{N}\mathcal{M}^4 e^{-\Gamma_\text{QSEA}[\bp_F]} .
\end{align}
Therefore, inserting our expressions for numerator and denominator into \Eq{eq:rate}, we find
\begin{align}
    \gamma_\text{FV} = \lim_{T \rightarrow \infty} \left\lvert 2 \, \text{Im} \, \left(\frac{1}{r} \frac{\frac{1}{2} \mathcal{J}[\bp_b] e^{-\Gamma_\text{QSEA}[\bp_b]} + \mathcal{J}[\bp_s] e^{-\Gamma_\text{QSEA}[\bp_s]}}{\mathcal{M}^4 \, e^{-\Gamma_\text{QSEA}[\bp_F]}}\right)_{\substack{r >0 \\ r =i}}\right\rvert .
\end{align}
It now remains to perform the analytic continuation $\T = i T$, or equivalently $r = i$.
The term $\mathcal{M}^4 = [U''(\phi_F)]^2$ doesn't rotate under the analytic continuation; nor, as we have currently defined it, does $\mathcal{J}$ (any phase $r$ introduced in $\mathcal{J}$ is cancelled by the integral measure; see discussion below Eq.~\eqref{eq:collective}).

Therefore, the two terms that are relevant are the factor of $1/r = -i$ and the exponential terms $e^{-\Gamma_\text{QSEA}}$, which we will now tackle. Just as in Eqs.~(4.23--4.28) in Ref.~\cite{Andreassen:2016cvx}, the action can be separated into a constant FV energy term proportional to time and a dynamic term not proportional to time,
\begin{align}
    \Gamma_\text{QSEA}[\bp_b] &= \T E_\text{FV} + \Gamma_b^0 ,
    \\
    \Gamma_\text{QSEA}[\bp_s] &= \T E_\text{TV} + \Gamma_s^0 ,
    \\
    \Gamma_\text{QSEA}[\bp_F] &= \T E_\text{FV} .
\end{align}
Therefore, when analytically continued from real $\T$ to imaginary $\T = i T$, we expect to find
\begin{align}
    \Gamma_\text{QSEA}[\bp_b] &\rightarrow i T E_\text{FV} + \Gamma_b^0 ,
    \\
    \Gamma_\text{QSEA}[\bp_s] &\rightarrow i T E_\text{TV} + \Gamma_s^0 ,
    \\
    \Gamma_\text{QSEA}[\bp_F] &\rightarrow i T E_\text{FV} .
\end{align}
In other words, while the QSEA will in general pick up an imaginary part under the analytic continuation, the terms $\Gamma_b^0 = \Gamma[\bp_b] - \Gamma[\bp_F]$ and $\Gamma_s^0 = \Gamma[\bp_s] - \Gamma[\bp_T]$ will remain real. As the real part of of the shot is exponentially suppressed relative to the bounce, we may drop it. This is the same analysis performed by Ref.~\cite{Andreassen:2016cvx}.

In light of the above discussion, we will find that after the analytic continuation we have a non-zero imaginary part,
\begin{align}
    \gamma_\text{FV} &\simeq \lim_{T\rightarrow \infty} \left\lvert 2 \,\text{Im} \left(- \frac{i}{2} \frac{(\Gamma[\bp_b] - \Gamma[\bp_F])^2}{16 \pi^4 \mathcal{M}^4} e^{- (\Gamma_\text{QSEA}[\bp_b] - \Gamma_\text{QSEA}[\bp_F])}\right)\right\rvert
    \\
    &= \frac{(\Gamma[\bp_b] - \Gamma[\bp_F])^2}{16 \pi^4 \mathcal{M}^4} e^{- (\Gamma_\text{QSEA}[\bp_b] - \Gamma_\text{QSEA}[\bp_F])}\ .
\end{align}

\section{Derivation of the QSEA propagator}
\label{sec:propagator_new_terms}
In this section, we derive the propagator $G_k[\bp;p,q]$ including the new terms arising from functional dependence of $W[J,\bp]$ on $\bp$ in the QSEA.
For the purposes of this appendix only, we will use $\delta F[f]/\delta f(x)$ to indicate total functional derivatives and $\partial F[f]/ \partial f(x)$ or $^{(n,m)}$ to indicate partial functional derivatives; elsewhere in the paper we have only used partial functional derivatives.
The definition of the propagator is
\begin{align}
    G_k = W_k^{(2,0)}[J,\bp] = \langle \phi^2 \rangle - \bp^2.
\end{align}
Ordinarily, one would identify this as the total derivative $\delta \bp /\delta J$, which by the functional chain rule is the inverse of $\delta J/ \delta \bp$.
However, this inversion is only possible for total derivatives.
For the QSEA, with additional functional dependence of $W_k[J,\bp]$ on $\bp$, the total derivative of $\bp$ produces additional terms:
\begin{align}
    \frac{\delta \bp(p)}{\delta J(q)} = \frac{\delta W_k^{(1,0)}[J,\bp;p]}{\delta J(q)} = (W_k^{(2,0)})_{p,q} + \int_{p'} \frac{\partial^2 W_k}{\partial J(p) \partial \bp(p')} \frac{\delta \bp(p')}{\delta J(q)}.
\end{align}
Rearranging, we can then write
\begin{align}
    (W_k^{(2,0)})_{p,q} = \int_{p'} \left[\delta(p-p') - \frac{\partial^2 W_k}{\partial J(p) \partial \bp(p')}\right] \frac{\delta \bp(p')}{\delta J(q)},
\end{align}
where here we have used $\partial$ to indicate partial functional derivatives and $\delta$ to indicate total functional derivatives.
We observe that, compared to the typical case, we have an additional term $W_k^{(1,1)}$.
Since $\delta \bp / \delta J$ is a total derivative, we can invert it $\delta \bp / \delta J = [\delta J / \delta \bp]^{-1}$ as we would usually.
However, this $\delta J / \delta \bp$ also receives additional terms.
Looking at the expression for $J$ \eqref{eq:JQSEA}, the total derivative yields
\begin{align}
    \frac{\delta J(p)}{\delta \bp(q)} = \Gamma_k^{(2)} + W_k^{(2,0)} + \Delta S_k^{(2)} + \int_{p'} (W_k^{(1,1)})_{p,p'} \frac{\delta J(p')}{\delta \bp(q)}
    \\
    \implies \int_{p'} \left[\delta(p-p') - \frac{\partial^2 W_k}{\partial \bp(p) \partial J(p')}\right] \frac{\delta J(p')}{\delta \bp(q)} = \Gamma_k^{(2)} + W_k^{(2,0)} + \Delta S_k^{(2)}.
\end{align}
If we integrate in the inverse of the first term, we get
\begin{align}
    \frac{\delta J(p)}{\delta \bp(q)} &= \int_{p'} \left[\delta(p-p') - \frac{\partial^2 W_k}{\partial \bp(p) \partial J(p')}\right]^{-1} \left[\Gamma_k^{(2)} + W_k^{(2,0)} + \Delta S_k^{(2)}\right]_{p',q}.
\end{align}
And so, inverting both sides,
\begin{align}
    \frac{\delta \bp(p)}{\delta J(q)} = \left[\frac{\delta J(q)}{\delta \bp(p)}\right]^{-1} =  \int_{p'} \left[\Gamma_k^{(2)} + W_k^{(2,0)} + \Delta S_k^{(2)}\right]^{-1}_{p,p'} \left[\delta(p'-q) - \frac{\partial^2 W_k}{\partial \bp(p') \partial J(q)}\right],
\end{align}
which allows us to finally write
\begin{equation}
    \begin{split}
        (W_k^{(2,0)})_{p,q} = \int_{p',q'} \left[\delta(p-p') - \frac{\partial^2 W_k}{\partial J(p) \partial \bp(p')}\right]  & \left[\Gamma_k^{(2)} + W_k^{(2,0)} + \Delta S_k^{(2)}\right]^{-1}_{p',q'} 
        \\
        & \cdot \left[\delta(q'-q) - \frac{\partial^2 W_k}{\partial \bp(q') \partial J(q)}\right].
    \end{split}
\end{equation}
In the above, $W_k^{(1,1)}$ is
\begin{align}
    \frac{\partial^2 W_k}{\partial J(p) \partial \bp(p')} &= \frac{1}{2} \int_{q} \frac{\delta R_k[\bp;q]}{\delta \bp(p')} \left(\langle \phi(p) \phi(-q)\phi(q)\rangle - \bp(p) \langle \phi(-q)\phi(q)\rangle \right)
    \\
    &= - \frac{V'''(\bp)}{2} \frac{\delta(p')}{\V} \int_{q} (\tilde{k}^2 - q^2) \left(\langle \phi(p) \phi(-q)\phi(q)\rangle - \bp(p) \langle \phi(-q)\phi(q)\rangle \right).
\end{align}
Even with our modifications, the LPA can only generate momentum-conserving correlators, as the action including our choice of regulator is invariant under spatial translations: $\langle \phi(p) \phi(-q) \phi(q) \rangle \propto \delta(p - q + q) = \delta(p)$ and $\bp(p) \langle \phi(-q)\phi(q)\rangle \propto \delta(p)$.
Then overall we find $(W_k^{(1,1)})_{p, p'} \propto \delta(p) \delta(p')$.
As is shown in App.~\ref{sec:propagator_LPA}, $[\Gamma_k^{(2)} + W_k^{(2,0)} + \Delta S_k^{(2)}]^{-1}_{p,q}$ only contains $\delta(p + q)$ and $\delta(p) \delta(q)$ terms.
So, when integrated with the $(\mathbbm{1} - W_k^{(1,1)})$ factors, all the $W_k^{(1,1)}$ terms have momentum dependence $\propto \delta(p) \delta(q)$.
Separating these out, the propagator is then
\begin{align}
    G_k = \left[\Gamma_k^{(2)} + W_k^{(2,0)} + \Delta S_k^{(2)}\right]^{-1}_{p,q} + \{\text{terms } \propto \delta(p) \delta(q)\}.
\end{align}
As mentioned, $[\Gamma_k^{(2)} + W_k^{(2,0)} + \Delta S_k^{(2)}]^{-1}_{p,q}$ will yield more $\delta(p) \delta(q)$ terms and, crucially, a piece $\propto \delta(p-q)$.
In the flow equation, the propagator only appears multiplied by $\partial_k R_k$. 
Hence, by choosing the regulator to vanish for the point $p = 0$ we may trivially exclude all $\delta(p) \delta(q)$ contributions, leaving only the remaining $\delta(p-q)$ component.
This piece is derived in detail in App.~\ref{sec:propagator_LPA}.

\section{The QSEA propagator in the LPA}\label{sec:propagator_LPA}
In the LPA, the equation for the propagator \eqref{eq:propagator} yields a closed-form analytic solution, as we will demonstrate here.
We will begin by deriving the propagator $G_k[\bp;p,q]$ by expanding the different pieces in \eqref{eq:propagator}; this step does not depend on the choice of approximation scheme.
First, the term $\Delta S_k^{(2)}$ gives
\begin{equation}
    \begin{split}
        \Delta S_k^{(2)}[\bp;p,q]
        &= R_k[\bp;p]\delta(p+q) + \left(\bp(p)\frac{\delta R_k[\bp;p]}{\delta \bp(q)} + p \leftrightarrow q\right) 
        \\
        &+ \frac{1}{2}\int_{p'} \frac{\delta^2 R_k[\bp;p']}{\delta \bp(q) \delta \bp(p)} \bp(-p')\bp(p') .
    \end{split}
\end{equation}
To expand the term $W_k^{(0,2)}[J,\bp]$, we begin with the first partial functional derivative
\begin{align}
    \frac{\delta W_k}{\delta \bp(p)} &= e^{- W_k} \frac{\delta}{\delta \bp(p)} e^{W_k}
    = - \frac{1}{2} \int_{p'} \frac{\delta R_k[\bp;p']}{\delta \bp(p)} \langle \phi(-p') \phi(p') \rangle .
\end{align}
Then, noting that
\begin{align}
    \frac{\delta}{\delta \bp(q)} \langle \phi(-p') \phi(p') \rangle &= \frac{\delta}{\delta \bp(q)} \frac{\int \D \phi \, e^{- S_k[\phi] +\int J \phi} \phi(-p) \phi(p)}{\int \D \phi \, e^{- S_k[\phi] +\int J \phi}}
    \\
    &\hspace{-7em} = -\frac{1}{2} \int_{q'} \frac{\delta R_k[\bp;q']}{\delta \bp(q')} \left[\langle \phi(-q') \phi(q') \phi(-p') \phi(p')\rangle - \langle \phi(-q') \phi(q')\rangle \langle \phi(-p') \phi(p')\rangle\right],
\end{align}
we find
\begin{equation}
\begin{split}
    (W_k^{(0,2)})_{p,q} =\,& - \frac{1}{2} \int_{p'} \frac{\delta^2 R_k[\bp;p']}{\delta \bp(p) \bp(q)} \left(W_k^{(2,0)} + \bp\bp\right)_{-p',p'}
    \\
    &- \frac{1}{4} \int_{p',q'} \frac{\delta R_k[\bp;p']}{\delta \bp(p)} \frac{\delta R_k[\bp;q']}{\delta \bp(q)} \left(\langle \phi^4 \rangle_{-p',p',-q',q'} - \langle \phi^2 \rangle_{-p',p'}\langle \phi^2 \rangle_{-q',q'}\right) .
\end{split}
\end{equation}
Putting it all together, we find that the $\int_q R_k^{(2)} \bp^2$ terms cancel, leaving
\begin{equation}
    \begin{split}
        G_k[\bp;p,q]^{-1} &= \Gamma_k^{(2)}[\bp;p,q] + R_k[\bp;p]\delta(p+q)
        \\
        &
        + \left(\bp(p)\frac{\delta R_k[\bp;p]}{\delta \bp(q)} + p \leftrightarrow q\right)
        - \frac{1}{2} \int_{p'} \frac{\delta^2 R_k[\bp;p']}{\delta \bp(p) \bp(q)} W_k^{(2,0)}
        \\
        &
        + \frac{1}{4} \int_{p',q'} \frac{\delta R_k[\bp;p']}{\delta \bp(p)} \frac{\delta R_k[\bp;q']}{\delta \bp(q)} (\langle \phi^4\rangle - \langle \phi^2 \rangle \langle \phi^2 \rangle)_{p',q'}\ .
    \end{split}
\end{equation}
Identifying $G_k[\bp;-p',p'] = (W_k^{(2,0)})_{-p',p'}$ and collecting the terms 
\begin{align}
    \Delta \equiv \left(\bp(p)\frac{\delta R_k[\bp;p]}{\delta \bp(q)} + p \leftrightarrow q\right)  + \frac{1}{4} \int_{p',q'} \frac{\delta R_k[\bp;p']}{\delta \bp(p)} \frac{\delta R_k[\bp;q']}{\delta \bp(q)} (\langle \phi^4\rangle - \langle \phi^2 \rangle \langle \phi^2 \rangle)_{p',q'}\ ,
\end{align}
we can write the above as
\begin{align}
    G_k[\bp;p,q]^{-1} =\,& \Gamma_k^{(2)}[\bp;p,q] + R_k[\bp;p]\delta(p+q)
    - \frac{1}{2} \int_{p'} \frac{\delta^2 R_k[\bp;p']}{\delta \bp(p) \bp(q)} G_k[\bp;-p',p']
        + \Delta .
\end{align}
In order to determine the flow equation for the effective potential within the local potential approximation, we evaluate both sides of the flow equation for constant field $\bp(x) = \bp$.
We reassure the reader that evaluating the flow equation at constant field is merely a tool to isolate the flow equation for $U_k$ in the DE, as $U_k$ is a function rather than functional of the field.
We will now evaluate each of the individual pieces within the propagator for the constant-field case.

As in the unmodified FRG, the second derivative of the action gives
\begin{align}
    \Gamma_k^{(2)}[\bp;p,q] &= \frac{\delta^2 \int_x \left[(\partial \bp(x))^2 + U_k(\bp(x))\right]}{\delta \phi(p) \delta \phi(q)}\Big\rvert_{\bp(x) = \bp} = (p^2 + U_k''(\bp)) \delta(p+q) .
\end{align}
The regulator $R_k$ is just given by \Eq{eq:regulator}; however, since the Heaviside function is enforced by the term $\partial_k R_k$ in the flow equation, it is automatically satisfied in the propagator and may be omitted.
When added to the second derivative of the action, the total result is then $p^2 + U_k'' + R_k = k^2 + U_k''(\bp) - V''(\bp)$.

The terms in $\Delta$ only contribute pieces proportional to delta functions of the propagator momentum $p$, which we will now demonstrate using to the form of the first derivative of the regulator function
\begin{align}
    \frac{\delta R_k[\bp;p']}{\delta \bp(p)} &= \frac{\delta}{\delta \bp(p)} \left(k^2 - \frac{1}{\V} \int_x [-\partial^2 + V''(\bp(x))]\right) \Theta\left(k^2 - \frac{1}{\V} \int_x [-\partial^2 + V''(\bp(x))]\right)
    \\
    &= - \frac{1}{\V} \int_x e^{i p x} V'''(\bp(x)) \Theta\left(k^2 - \frac{1}{\V} \int_x [-\partial^2 + V''(\bp(x))]\right)
    \\
    & = - V'''(\bp) \Theta\left(k^2 - (p')^2 - V''(\bp)\right) \frac{\delta(p)}{\V} 
    \quad \quad (\text{constant } \bp(x) = \bp).
\end{align}
Again, the Heaviside function is automatically satisfied within the flow equation. Likewise, at constant position space field, the momentum space field is $\bp(p) = \bp \, \delta(p)$. Then, writing $\tilde{k}^2(\phi) \equiv k^2 - V''(\phi)$, the term $\Delta$ gives 
\begin{align}
        \Delta_{p,q} = \delta(p) \delta(q) \Bigg[&- 2
        \frac{\bp V'''(\bp)}{\V} \Theta(\tilde{k}^2) 
        \\
        & + \frac{V'''(\bp)^2 }{4 \V^2} \int_{p',q'} \Theta(\tilde{k}^2 - p'^2 ) \Theta(\tilde{k}^2 - q'^2 ) \left(\langle \phi^4 \rangle - \langle\phi^2\rangle \langle \phi^2 \rangle\right) \Bigg] .
\end{align}
Lastly, the second derivative of the regulator yields
\begin{align}
    \frac{\delta^2 R_k[\bp;p']}{\delta \bp(p) \delta \bp(q)} &= - \frac{\delta}{\delta \bp(q)} \frac{1}{\V} \int_x e^{i p x} V'''(\bp(x)) \Theta\left(k^2 - \frac{1}{\V} \int_x [-\partial^2 + V''(\bp(x))]\right)
    \\
    &= -\frac{1}{\V} \int e^{- i (p+q) x} V''''(\bp(x)) \Theta\left(k^2 - \frac{1}{\V} \int_x [-\partial^2 + V''(\bp(x))]\right)
    \\
    &+ \frac{1}{\V^2} \int_{x,y} e^{i (p x + q y)} V'''(\bp(x)) V'''(\bp(y)) \,\, \delta\left(k^2 - \frac{1}{\V} \int_x [-\partial^2 + V''(\bp(x))]\right)
    \\
    &= - V''''(\bp) \frac{\delta(p+q)}{\V} 
    \Theta(\tilde{k}^2 - p'^2)
    + V'''(\bp)^2 \frac{\delta(p) \delta(q)}{\V^2} \delta (\tilde{k}^2 - p'^2)
    \\
    &\hspace{20em} (\text{constant } \bp(x) = \bp ) .\nonumber
\end{align}
In summary, within the flow equation in the LPA, the propagator yields
\begin{equation}
    \begin{split}
        G_k[\bp;p,q]^{-1} &= \delta(p+q) \left[\tilde{k}^2 + U_k'' + \frac{1}{2} \int_{p'} V'''' \frac{G_k[\bp;-p',p']}{\V} \Theta(\tilde{k}^2 - p'^2)\right] 
        \\
        & + \left\{\text{terms}\propto \delta(p) \delta(q)\right\} .
    \end{split}
\end{equation}
Identifying $G_k(\phi)$ \eqref{eq:Gphi}, we can summarize this as
\begin{align}
    G_k[\bp;p,q]^{-1} = \frac{\delta(p+q)}{G_k(\phi)} + B \, \delta(p) \delta(q)
\end{align}
for some terms $B$.
We can now invert the propagator. Taking an ansatz for the propagator $G_k[\bp;p,q] = C \delta(p+q) + D \delta(p)\delta(q)$,
\begin{align}
    &\int_{p'} (C\delta(p+p') + D\delta(p) \delta(p')) \left(\frac{\delta(p'+q)}{G_k(\phi)} + B \delta(p') \delta(q)\right)
    \\
    &= C G_k(\phi) \delta(p-q) + \left(D G_k(\phi) + C B + \V B D\right) \delta(p) \delta(q) \stackrel{!}{=} \delta(p-q).
\end{align}
And so we require $C G_k(\phi) = 1$ and $D G_k(\phi) + C B + \V B D = 0$, so
\begin{align}
    G_k[\bp;p,q] = G_k(\phi) \delta(p+q) - \frac{G_k(\phi) B}{(1/G_k(\phi) + \V B)} \delta(p)\delta(q) .
\end{align}
We observe that since $G_k(\phi) > 0$, the coefficient $\delta(p)\delta(q)$ term is strictly $< G_k(\phi)/\V$, so that
\begin{align}
    G_k[\bp;p,q] = G_k(\phi) \delta(p+q) - \mathcal{O}\left(\frac{G_k(\phi)}{\V}\right) \delta(p)\delta(q).
\end{align}
As discussed in App.~\ref{sec:propagator_new_terms}, the $\delta(p) \delta(q)$ terms may be dropped by choosing the regulator to vanish at zero momentum.
Even if left in, they are ultimately suppressed by a factor of $1/(\V \tilde{k}^4)$ and so vanish in the large volume limit.
Within the context of the flow equation \eqref{eq:floweq} and the denominator of $G_k(\phi)$ \eqref{eq:Gphi}, we have the integral
\begin{align}
    \int_p \frac{G_k[\bp;-p,p]}{\V} \Theta(\tilde{k}^2 - p^2) = \int_p \frac{1}{\V} G_k(\phi) \delta(0) \Theta(\tilde{k}^2 - p^2)
    = \frac{\tilde{k}^4}{32 \pi^2} G_k(\phi)
\end{align}
Evaluating the integral in the equation for $G_k(\phi)$ \eqref{eq:Gphi} as above, we find an equation \eqref{eq:GphiInt} with closed-form solution \eqref{eq:closed_form}.
This closed-form solution can then be inserted into the flow equation.
%
%
%
%
Evaluating the integral in the equation for $G_k(\phi)$ \eqref{eq:Gphi} as above, we find an equation \eqref{eq:GphiInt} with closed-form solution \eqref{eq:closed_form}.
This closed-form solution can then be inserted into the flow equation.

\section{Worked example}

Within the LPA, the quasi-stationary effective potential may be found by solving the flow equation \eqref{eq:flowLPA} for $U_{k=0}$ with the initial value $U_{k=\Lambda} = V(\phi)$.
As the flow equation is an ordinary differential equation over $k$, it can be solved using preexisting differential equation solvers.
For the purposes of this paper, we discretized $U_k(\phi)$ over 100 points in $\phi \in (-0.25,1.25)$ and evaluated $U_k''(\phi)$ using second-order finite differences.
The flow equation in $k$ was then solved from the boundary $k = \Lambda$ to the final value $k = 0$ using SciPy's built-in differential equation solver \texttt{integrate.solve\_ivp}, using the \texttt{RK45} fifth/fourth-order explicit Runge-Kutta method.
The resulting quasi-stationary effective potential $U_{k=0}$ is shown in solid black in \Fig{fig:toymodel}.

For comparison, the analogous one-loop potential may be derived from the definition of the 1PI effective action
\begin{align}
    \Gamma[\bp] = -\ln \int D \phi e^{-S[\phi] + \int J \phi} + \int J \bp
\end{align}
which, when expanded at Gaussian order around $\bp$, yields
\begin{align}
    \Gamma[\bp] &\simeq  -\ln \int D \phi \, e^{-S[\bp] - \frac{1}{2} \int S''[\bp] (\phi- \bp)^2 + \int (J - S'[\bp]) (\phi-\bp)}
\end{align}
Since $J = \Gamma^{(1)} \simeq S^{(1)}$, at leading order the Gaussian integral yields
\begin{align}
    \Gamma[\bp] = S[\bp] + \frac{1}{2} \ln \det S''[\bp] = S[\bp] + \frac{1}{2} \int_p \ln \left[p^2 + V''(\bp)\right]
\end{align}
Renormalizing against the free theory yields the traditional expression for the one-loop effective potential.
\begin{align}
    V_\text{eff}(\phi) = V(\phi) + \frac{1}{2} \int_p \ln \left[\frac{p^2 + V''(\phi)}{p^2 + m^2}\right]
\end{align}
in terms of the field-dependent mass $V''(\phi) = m^2 + \alpha \phi + \lambda \phi^2/2$. Then, performing the integral with a momentum cutoff $p_\text{max}$ related to the QSEA fluctuation cutoff $\Lambda$ as $p_\text{max}^2 = \Lambda^2 - V''(\phi)$, we find that the explicit expression for the one-loop effective potential is
\begin{align}\label{eq:veff_oneloop}
    V_\text{eff} = V 
    + \frac{1}{64 \pi^2} \bigg(m^4 \ln \left[\frac{m^2 + p_\text{max}^2}{m^2}\right] - (V'')^2 \ln \left[\frac{V'' + p_\text{max}^2}{V''}\right]& 
    \\
    + p_\text{max}^4 \ln \left[\frac{V'' + p_\text{max}^2}{m^2 + p_\text{max}^2}\right] + p_\text{max}^2 \left(V''- m^2\right)&\bigg).
\end{align}
$V(\phi)$ should be considered the bare potential with fluctuation cutoff $\Lambda$ (momentum cutoff $p_\text{max}$), and so we do not need to consider the renormalization of the effective potential for the purposes of this paper.
The effective potential \eqref{eq:veff_oneloop} for the toy model presented in the paper is plotted in \Fig{fig:toymodel} as a red dashed line.
As expected, the one-loop result agrees with the QSEA in the weak-coupling limit but begins to differ in the strong-coupling limit.

Lastly, we also contrast our result with the unmodified functional renormalization group in the LPA with Litim's optimized regulator $R_k = (k^2 - p^2) \Theta(k^2 - p^2)$.
The (momentum-space) flow equation in this case can be evaluated to give
\begin{align}
    \partial_k U_k &= \frac{1}{2 \V} \int_p \left[\Gamma_k^{(2)}[\bp] + R_k \delta(p+q)\right]^{-1} \partial_k R_k 
    \\
    &=\frac{1}{8 \pi^2} \int_0^{p_\text{max}} \td p\, p^3 \frac{k \Theta(k^2 - p^2)}{k^2 + U_k''}
    \\
    &= \frac{k \, \min\{k^4,p_\text{max}^4\}}{32 \pi^2} \frac{1}{k^2 + U_k''}\ .
\end{align}
Taking an FRG cutoff $\Lambda_\text{FRG} = p_\text{max}$ to match the QSEA cutoff, we get that the minimum always yields $k^4$ and so the flow equation reduces to the well-known result
\begin{equation}
    \partial_k U_k = \frac{k^5}{32 \pi^2} \frac{1}{k^2 + U_k''} \Theta(p_\text{max}^2 - k^2)
\end{equation}
with an additional condition $\Theta(p_\text{max}^2 - k^2)$ due to the nature of the fluctuation cutoff.
The flow equation for the unmodified FRG can be evaluated analogously to the QSEA; however, due to the stiffness of the unmodified FRG from the pole $1/(k^2 + U_k'')$, we used the implicit \texttt{BDF} solution method instead of \texttt{RK45}. 
All results have been independently verified with a method-of-lines calculation at second order finite differences for the discrete dimension (the field $\bar\phi$) in Wolfram mathematica.
The result from the unmodified FRG is plotted in dotted violet in \Fig{fig:toymodel}.
As expected, the unmodified FRG is convex and cannot describe tunneling; however, it does correctly reproduce the location of the quantum false vacuum.

As a technical note, we point out that in each method we observe the expected quadratic UV divergence of the mass.
With such a UV divergence, the Litim regulator no longer functions exactly as a hard momentum cutoff in the UV, as UV momenta still make a contribution to the running of the mass.
However, because the running of the mass becomes exactly quadratic in this limit and no other terms are similarly divergent, counterterms can be introduced to exactly subtract off any such difference.
While something to be aware of in generality, it turns out that with the QSEA cutoff use in this paper, the UV behavior is the same for perturbation theory and the unmodified FRG and is not a relevant issue.
There is a slight difference between the unmodified FRG and the QSEA flow owing to the correction to the flow equation, amounting to a difference in the flow with relative scale
\begin{align}
   1 - \frac{2}{1 + \sqrt{1 + \frac{V''''(\phi)}{16 \pi^2}}}
\end{align}
The extent to which this different UV flow is a true correction or something that should be removed with counterterms will be investigated in future works; we have not corrected for it in this work.
In practice, however, even in the strong coupling limit the resulting difference in the QSEA potential with and with out the counterterm is essentially negligible ($\sim 2\%$) compared to the difference between the QSEA and perturbation theory.

\bibliographystyle{apsrev4-1}
\bibliography{refs}

\begin{thebibliography}{67}%
\makeatletter
\providecommand \@ifxundefined [1]{%
 \@ifx{#1\undefined}
}%
\providecommand \@ifnum [1]{%
 \ifnum #1\expandafter \@firstoftwo
 \else \expandafter \@secondoftwo
 \fi
}%
\providecommand \@ifx [1]{%
 \ifx #1\expandafter \@firstoftwo
 \else \expandafter \@secondoftwo
 \fi
}%
\providecommand \natexlab [1]{#1}%
\providecommand \enquote  [1]{``#1''}%
\providecommand \bibnamefont  [1]{#1}%
\providecommand \bibfnamefont [1]{#1}%
\providecommand \citenamefont [1]{#1}%
\providecommand \href@noop [0]{\@secondoftwo}%
\providecommand \href [0]{\begingroup \@sanitize@url \@href}%
\providecommand \@href[1]{\@@startlink{#1}\@@href}%
\providecommand \@@href[1]{\endgroup#1\@@endlink}%
\providecommand \@sanitize@url [0]{\catcode `\\12\catcode `\$12\catcode
  `\&12\catcode `\#12\catcode `\^12\catcode `\_12\catcode `\%12\relax}%
\providecommand \@@startlink[1]{}%
\providecommand \@@endlink[0]{}%
\providecommand \url  [0]{\begingroup\@sanitize@url \@url }%
\providecommand \@url [1]{\endgroup\@href {#1}{\urlprefix }}%
\providecommand \urlprefix  [0]{URL }%
\providecommand \Eprint [0]{\href }%
\providecommand \doibase [0]{http://dx.doi.org/}%
\providecommand \selectlanguage [0]{\@gobble}%
\providecommand \bibinfo  [0]{\@secondoftwo}%
\providecommand \bibfield  [0]{\@secondoftwo}%
\providecommand \translation [1]{[#1]}%
\providecommand \BibitemOpen [0]{}%
\providecommand \bibitemStop [0]{}%
\providecommand \bibitemNoStop [0]{.\EOS\space}%
\providecommand \EOS [0]{\spacefactor3000\relax}%
\providecommand \BibitemShut  [1]{\csname bibitem#1\endcsname}%
\let\auto@bib@innerbib\@empty
\bibitem [{\citenamefont {Coleman}(1977)}]{Coleman:1977py}%
  \BibitemOpen
  \bibfield  {author} {\bibinfo {author} {\bibfnamefont {S.~R.}\ \bibnamefont
  {Coleman}},\ }\href {\doibase 10.1103/PhysRevD.16.1248} {\bibfield  {journal}
  {\bibinfo  {journal} {Phys. Rev. D}\ }\textbf {\bibinfo {volume} {15}},\
  \bibinfo {pages} {2929} (\bibinfo {year} {1977})},\ \bibinfo {note}
  {[Erratum: Phys.Rev.D 16, 1248(E) (1977)]}\BibitemShut {NoStop}%
\bibitem [{\citenamefont {Callan}\ and\ \citenamefont
  {Coleman}(1977)}]{Callan:1977pt}%
  \BibitemOpen
  \bibfield  {author} {\bibinfo {author} {\bibfnamefont {C.~G.}\ \bibnamefont
  {Callan}, \bibfnamefont {Jr.}}\ and\ \bibinfo {author} {\bibfnamefont
  {S.~R.}\ \bibnamefont {Coleman}},\ }\href {\doibase 10.1103/PhysRevD.16.1762}
  {\bibfield  {journal} {\bibinfo  {journal} {Phys. Rev. D}\ }\textbf {\bibinfo
  {volume} {16}},\ \bibinfo {pages} {1762} (\bibinfo {year}
  {1977})}\BibitemShut {NoStop}%
\bibitem [{\citenamefont {Weinberg}(1993)}]{Weinberg:1992ds}%
  \BibitemOpen
  \bibfield  {author} {\bibinfo {author} {\bibfnamefont {E.~J.}\ \bibnamefont
  {Weinberg}},\ }\href {\doibase 10.1103/PhysRevD.47.4614} {\bibfield
  {journal} {\bibinfo  {journal} {Phys. Rev. D}\ }\textbf {\bibinfo {volume}
  {47}},\ \bibinfo {pages} {4614} (\bibinfo {year} {1993})},\ \Eprint
  {http://arxiv.org/abs/hep-ph/9211314} {arXiv:hep-ph/9211314} \BibitemShut
  {NoStop}%
\bibitem [{\citenamefont {Langer}(1969)}]{Langer:1969bc}%
  \BibitemOpen
  \bibfield  {author} {\bibinfo {author} {\bibfnamefont {J.~S.}\ \bibnamefont
  {Langer}},\ }\href {\doibase 10.1016/0003-4916(69)90153-5} {\bibfield
  {journal} {\bibinfo  {journal} {Annals Phys.}\ }\textbf {\bibinfo {volume}
  {54}},\ \bibinfo {pages} {258} (\bibinfo {year} {1969})}\BibitemShut
  {NoStop}%
\bibitem [{\citenamefont {Wetterich}(1993)}]{Wetterich:1992yh}%
  \BibitemOpen
  \bibfield  {author} {\bibinfo {author} {\bibfnamefont {C.}~\bibnamefont
  {Wetterich}},\ }\href {\doibase 10.1016/0370-2693(93)90726-X} {\bibfield
  {journal} {\bibinfo  {journal} {Phys. Lett. B}\ }\textbf {\bibinfo {volume}
  {301}},\ \bibinfo {pages} {90} (\bibinfo {year} {1993})},\ \Eprint
  {http://arxiv.org/abs/1710.05815} {arXiv:1710.05815 [hep-th]} \BibitemShut
  {NoStop}%
\bibitem [{\citenamefont {Braun}(2009)}]{Braun:2008pi}%
  \BibitemOpen
  \bibfield  {author} {\bibinfo {author} {\bibfnamefont {J.}~\bibnamefont
  {Braun}},\ }\href {\doibase 10.1140/epjc/s10052-009-1136-6} {\bibfield
  {journal} {\bibinfo  {journal} {Eur. Phys. J. C}\ }\textbf {\bibinfo {volume}
  {64}},\ \bibinfo {pages} {459} (\bibinfo {year} {2009})},\ \Eprint
  {http://arxiv.org/abs/0810.1727} {arXiv:0810.1727 [hep-ph]} \BibitemShut
  {NoStop}%
\bibitem [{\citenamefont {Braun}\ \emph {et~al.}(2011)\citenamefont {Braun},
  \citenamefont {Haas}, \citenamefont {Marhauser},\ and\ \citenamefont
  {Pawlowski}}]{Braun:2009gm}%
  \BibitemOpen
  \bibfield  {author} {\bibinfo {author} {\bibfnamefont {J.}~\bibnamefont
  {Braun}}, \bibinfo {author} {\bibfnamefont {L.~M.}\ \bibnamefont {Haas}},
  \bibinfo {author} {\bibfnamefont {F.}~\bibnamefont {Marhauser}}, \ and\
  \bibinfo {author} {\bibfnamefont {J.~M.}\ \bibnamefont {Pawlowski}},\ }\href
  {\doibase 10.1103/PhysRevLett.106.022002} {\bibfield  {journal} {\bibinfo
  {journal} {Phys. Rev. Lett.}\ }\textbf {\bibinfo {volume} {106}},\ \bibinfo
  {pages} {022002} (\bibinfo {year} {2011})},\ \Eprint
  {http://arxiv.org/abs/0908.0008} {arXiv:0908.0008 [hep-ph]} \BibitemShut
  {NoStop}%
\bibitem [{\citenamefont {Skokov}\ \emph {et~al.}(2010)\citenamefont {Skokov},
  \citenamefont {Stokic}, \citenamefont {Friman},\ and\ \citenamefont
  {Redlich}}]{Skokov:2010wb}%
  \BibitemOpen
  \bibfield  {author} {\bibinfo {author} {\bibfnamefont {V.}~\bibnamefont
  {Skokov}}, \bibinfo {author} {\bibfnamefont {B.}~\bibnamefont {Stokic}},
  \bibinfo {author} {\bibfnamefont {B.}~\bibnamefont {Friman}}, \ and\ \bibinfo
  {author} {\bibfnamefont {K.}~\bibnamefont {Redlich}},\ }\href {\doibase
  10.1103/PhysRevC.82.015206} {\bibfield  {journal} {\bibinfo  {journal} {Phys.
  Rev. C}\ }\textbf {\bibinfo {volume} {82}},\ \bibinfo {pages} {015206}
  (\bibinfo {year} {2010})},\ \Eprint {http://arxiv.org/abs/1004.2665}
  {arXiv:1004.2665 [hep-ph]} \BibitemShut {NoStop}%
\bibitem [{\citenamefont {Herbst}\ \emph {et~al.}(2011)\citenamefont {Herbst},
  \citenamefont {Pawlowski},\ and\ \citenamefont {Schaefer}}]{Herbst:2010rf}%
  \BibitemOpen
  \bibfield  {author} {\bibinfo {author} {\bibfnamefont {T.~K.}\ \bibnamefont
  {Herbst}}, \bibinfo {author} {\bibfnamefont {J.~M.}\ \bibnamefont
  {Pawlowski}}, \ and\ \bibinfo {author} {\bibfnamefont {B.-J.}\ \bibnamefont
  {Schaefer}},\ }\href {\doibase 10.1016/j.physletb.2010.12.003} {\bibfield
  {journal} {\bibinfo  {journal} {Phys. Lett. B}\ }\textbf {\bibinfo {volume}
  {696}},\ \bibinfo {pages} {58} (\bibinfo {year} {2011})},\ \Eprint
  {http://arxiv.org/abs/1008.0081} {arXiv:1008.0081 [hep-ph]} \BibitemShut
  {NoStop}%
\bibitem [{\citenamefont {Skokov}\ \emph {et~al.}(2011)\citenamefont {Skokov},
  \citenamefont {Friman},\ and\ \citenamefont {Redlich}}]{Skokov:2010uh}%
  \BibitemOpen
  \bibfield  {author} {\bibinfo {author} {\bibfnamefont {V.}~\bibnamefont
  {Skokov}}, \bibinfo {author} {\bibfnamefont {B.}~\bibnamefont {Friman}}, \
  and\ \bibinfo {author} {\bibfnamefont {K.}~\bibnamefont {Redlich}},\ }\href
  {\doibase 10.1103/PhysRevC.83.054904} {\bibfield  {journal} {\bibinfo
  {journal} {Phys. Rev. C}\ }\textbf {\bibinfo {volume} {83}},\ \bibinfo
  {pages} {054904} (\bibinfo {year} {2011})},\ \Eprint
  {http://arxiv.org/abs/1008.4570} {arXiv:1008.4570 [hep-ph]} \BibitemShut
  {NoStop}%
\bibitem [{\citenamefont {Fister}\ and\ \citenamefont
  {Pawlowski}(2011)}]{Fister:2011uw}%
  \BibitemOpen
  \bibfield  {author} {\bibinfo {author} {\bibfnamefont {L.}~\bibnamefont
  {Fister}}\ and\ \bibinfo {author} {\bibfnamefont {J.~M.}\ \bibnamefont
  {Pawlowski}},\ }\href@noop {} {\  (\bibinfo {year} {2011})},\ \Eprint
  {http://arxiv.org/abs/1112.5440} {arXiv:1112.5440 [hep-ph]} \BibitemShut
  {NoStop}%
\bibitem [{\citenamefont {Herbst}\ \emph {et~al.}(2013)\citenamefont {Herbst},
  \citenamefont {Pawlowski},\ and\ \citenamefont {Schaefer}}]{Herbst:2013ail}%
  \BibitemOpen
  \bibfield  {author} {\bibinfo {author} {\bibfnamefont {T.~K.}\ \bibnamefont
  {Herbst}}, \bibinfo {author} {\bibfnamefont {J.~M.}\ \bibnamefont
  {Pawlowski}}, \ and\ \bibinfo {author} {\bibfnamefont {B.-J.}\ \bibnamefont
  {Schaefer}},\ }\href {\doibase 10.1103/PhysRevD.88.014007} {\bibfield
  {journal} {\bibinfo  {journal} {Phys. Rev. D}\ }\textbf {\bibinfo {volume}
  {88}},\ \bibinfo {pages} {014007} (\bibinfo {year} {2013})},\ \Eprint
  {http://arxiv.org/abs/1302.1426} {arXiv:1302.1426 [hep-ph]} \BibitemShut
  {NoStop}%
\bibitem [{\citenamefont {Mitter}\ \emph {et~al.}(2015)\citenamefont {Mitter},
  \citenamefont {Pawlowski},\ and\ \citenamefont
  {Strodthoff}}]{Mitter:2014wpa}%
  \BibitemOpen
  \bibfield  {author} {\bibinfo {author} {\bibfnamefont {M.}~\bibnamefont
  {Mitter}}, \bibinfo {author} {\bibfnamefont {J.~M.}\ \bibnamefont
  {Pawlowski}}, \ and\ \bibinfo {author} {\bibfnamefont {N.}~\bibnamefont
  {Strodthoff}},\ }\href {\doibase 10.1103/PhysRevD.91.054035} {\bibfield
  {journal} {\bibinfo  {journal} {Phys. Rev. D}\ }\textbf {\bibinfo {volume}
  {91}},\ \bibinfo {pages} {054035} (\bibinfo {year} {2015})},\ \Eprint
  {http://arxiv.org/abs/1411.7978} {arXiv:1411.7978 [hep-ph]} \BibitemShut
  {NoStop}%
\bibitem [{\citenamefont {Braun}\ \emph {et~al.}(2016)\citenamefont {Braun},
  \citenamefont {Fister}, \citenamefont {Pawlowski},\ and\ \citenamefont
  {Rennecke}}]{Braun:2014ata}%
  \BibitemOpen
  \bibfield  {author} {\bibinfo {author} {\bibfnamefont {J.}~\bibnamefont
  {Braun}}, \bibinfo {author} {\bibfnamefont {L.}~\bibnamefont {Fister}},
  \bibinfo {author} {\bibfnamefont {J.~M.}\ \bibnamefont {Pawlowski}}, \ and\
  \bibinfo {author} {\bibfnamefont {F.}~\bibnamefont {Rennecke}},\ }\href
  {\doibase 10.1103/PhysRevD.94.034016} {\bibfield  {journal} {\bibinfo
  {journal} {Phys. Rev. D}\ }\textbf {\bibinfo {volume} {94}},\ \bibinfo
  {pages} {034016} (\bibinfo {year} {2016})},\ \Eprint
  {http://arxiv.org/abs/1412.1045} {arXiv:1412.1045 [hep-ph]} \BibitemShut
  {NoStop}%
\bibitem [{\citenamefont {Rennecke}(2015)}]{Rennecke:2015eba}%
  \BibitemOpen
  \bibfield  {author} {\bibinfo {author} {\bibfnamefont {F.}~\bibnamefont
  {Rennecke}},\ }\href {\doibase 10.1103/PhysRevD.92.076012} {\bibfield
  {journal} {\bibinfo  {journal} {Phys. Rev. D}\ }\textbf {\bibinfo {volume}
  {92}},\ \bibinfo {pages} {076012} (\bibinfo {year} {2015})},\ \Eprint
  {http://arxiv.org/abs/1504.03585} {arXiv:1504.03585 [hep-ph]} \BibitemShut
  {NoStop}%
\bibitem [{\citenamefont {Cyrol}\ \emph {et~al.}(2016)\citenamefont {Cyrol},
  \citenamefont {Fister}, \citenamefont {Mitter}, \citenamefont {Pawlowski},\
  and\ \citenamefont {Strodthoff}}]{Cyrol:2016tym}%
  \BibitemOpen
  \bibfield  {author} {\bibinfo {author} {\bibfnamefont {A.~K.}\ \bibnamefont
  {Cyrol}}, \bibinfo {author} {\bibfnamefont {L.}~\bibnamefont {Fister}},
  \bibinfo {author} {\bibfnamefont {M.}~\bibnamefont {Mitter}}, \bibinfo
  {author} {\bibfnamefont {J.~M.}\ \bibnamefont {Pawlowski}}, \ and\ \bibinfo
  {author} {\bibfnamefont {N.}~\bibnamefont {Strodthoff}},\ }\href {\doibase
  10.1103/PhysRevD.94.054005} {\bibfield  {journal} {\bibinfo  {journal} {Phys.
  Rev. D}\ }\textbf {\bibinfo {volume} {94}},\ \bibinfo {pages} {054005}
  (\bibinfo {year} {2016})},\ \Eprint {http://arxiv.org/abs/1605.01856}
  {arXiv:1605.01856 [hep-ph]} \BibitemShut {NoStop}%
\bibitem [{\citenamefont {Fu}\ \emph {et~al.}(2016)\citenamefont {Fu},
  \citenamefont {Pawlowski}, \citenamefont {Rennecke},\ and\ \citenamefont
  {Schaefer}}]{Fu:2016tey}%
  \BibitemOpen
  \bibfield  {author} {\bibinfo {author} {\bibfnamefont {W.-j.}\ \bibnamefont
  {Fu}}, \bibinfo {author} {\bibfnamefont {J.~M.}\ \bibnamefont {Pawlowski}},
  \bibinfo {author} {\bibfnamefont {F.}~\bibnamefont {Rennecke}}, \ and\
  \bibinfo {author} {\bibfnamefont {B.-J.}\ \bibnamefont {Schaefer}},\ }\href
  {\doibase 10.1103/PhysRevD.94.116020} {\bibfield  {journal} {\bibinfo
  {journal} {Phys. Rev. D}\ }\textbf {\bibinfo {volume} {94}},\ \bibinfo
  {pages} {116020} (\bibinfo {year} {2016})},\ \Eprint
  {http://arxiv.org/abs/1608.04302} {arXiv:1608.04302 [hep-ph]} \BibitemShut
  {NoStop}%
\bibitem [{\citenamefont {Cyrol}\ \emph
  {et~al.}(2018{\natexlab{a}})\citenamefont {Cyrol}, \citenamefont {Mitter},
  \citenamefont {Pawlowski},\ and\ \citenamefont {Strodthoff}}]{Cyrol:2017ewj}%
  \BibitemOpen
  \bibfield  {author} {\bibinfo {author} {\bibfnamefont {A.~K.}\ \bibnamefont
  {Cyrol}}, \bibinfo {author} {\bibfnamefont {M.}~\bibnamefont {Mitter}},
  \bibinfo {author} {\bibfnamefont {J.~M.}\ \bibnamefont {Pawlowski}}, \ and\
  \bibinfo {author} {\bibfnamefont {N.}~\bibnamefont {Strodthoff}},\ }\href
  {\doibase 10.1103/PhysRevD.97.054006} {\bibfield  {journal} {\bibinfo
  {journal} {Phys. Rev. D}\ }\textbf {\bibinfo {volume} {97}},\ \bibinfo
  {pages} {054006} (\bibinfo {year} {2018}{\natexlab{a}})},\ \Eprint
  {http://arxiv.org/abs/1706.06326} {arXiv:1706.06326 [hep-ph]} \BibitemShut
  {NoStop}%
\bibitem [{\citenamefont {Cyrol}\ \emph
  {et~al.}(2018{\natexlab{b}})\citenamefont {Cyrol}, \citenamefont {Mitter},
  \citenamefont {Pawlowski},\ and\ \citenamefont {Strodthoff}}]{Cyrol:2017qkl}%
  \BibitemOpen
  \bibfield  {author} {\bibinfo {author} {\bibfnamefont {A.~K.}\ \bibnamefont
  {Cyrol}}, \bibinfo {author} {\bibfnamefont {M.}~\bibnamefont {Mitter}},
  \bibinfo {author} {\bibfnamefont {J.~M.}\ \bibnamefont {Pawlowski}}, \ and\
  \bibinfo {author} {\bibfnamefont {N.}~\bibnamefont {Strodthoff}},\ }\href
  {\doibase 10.1103/PhysRevD.97.054015} {\bibfield  {journal} {\bibinfo
  {journal} {Phys. Rev. D}\ }\textbf {\bibinfo {volume} {97}},\ \bibinfo
  {pages} {054015} (\bibinfo {year} {2018}{\natexlab{b}})},\ \Eprint
  {http://arxiv.org/abs/1708.03482} {arXiv:1708.03482 [hep-ph]} \BibitemShut
  {NoStop}%
\bibitem [{\citenamefont {Fu}\ \emph {et~al.}(2020)\citenamefont {Fu},
  \citenamefont {Pawlowski},\ and\ \citenamefont {Rennecke}}]{Fu:2019hdw}%
  \BibitemOpen
  \bibfield  {author} {\bibinfo {author} {\bibfnamefont {W.-j.}\ \bibnamefont
  {Fu}}, \bibinfo {author} {\bibfnamefont {J.~M.}\ \bibnamefont {Pawlowski}}, \
  and\ \bibinfo {author} {\bibfnamefont {F.}~\bibnamefont {Rennecke}},\ }\href
  {\doibase 10.1103/PhysRevD.101.054032} {\bibfield  {journal} {\bibinfo
  {journal} {Phys. Rev. D}\ }\textbf {\bibinfo {volume} {101}},\ \bibinfo
  {pages} {054032} (\bibinfo {year} {2020})},\ \Eprint
  {http://arxiv.org/abs/1909.02991} {arXiv:1909.02991 [hep-ph]} \BibitemShut
  {NoStop}%
\bibitem [{\citenamefont {Dupuis}\ \emph {et~al.}(2021)\citenamefont {Dupuis},
  \citenamefont {Canet}, \citenamefont {Eichhorn}, \citenamefont {Metzner},
  \citenamefont {Pawlowski}, \citenamefont {Tissier},\ and\ \citenamefont
  {Wschebor}}]{Dupuis:2020fhh}%
  \BibitemOpen
  \bibfield  {author} {\bibinfo {author} {\bibfnamefont {N.}~\bibnamefont
  {Dupuis}}, \bibinfo {author} {\bibfnamefont {L.}~\bibnamefont {Canet}},
  \bibinfo {author} {\bibfnamefont {A.}~\bibnamefont {Eichhorn}}, \bibinfo
  {author} {\bibfnamefont {W.}~\bibnamefont {Metzner}}, \bibinfo {author}
  {\bibfnamefont {J.~M.}\ \bibnamefont {Pawlowski}}, \bibinfo {author}
  {\bibfnamefont {M.}~\bibnamefont {Tissier}}, \ and\ \bibinfo {author}
  {\bibfnamefont {N.}~\bibnamefont {Wschebor}},\ }\href {\doibase
  10.1016/j.physrep.2021.01.001} {\bibfield  {journal} {\bibinfo  {journal}
  {Phys. Rept.}\ }\textbf {\bibinfo {volume} {910}},\ \bibinfo {pages} {1}
  (\bibinfo {year} {2021})},\ \Eprint {http://arxiv.org/abs/2006.04853}
  {arXiv:2006.04853 [cond-mat.stat-mech]} \BibitemShut {NoStop}%
\bibitem [{\citenamefont {Witten}(2011)}]{Witten:2010cx}%
  \BibitemOpen
  \bibfield  {author} {\bibinfo {author} {\bibfnamefont {E.}~\bibnamefont
  {Witten}},\ }\href@noop {} {\bibfield  {journal} {\bibinfo  {journal} {AMS/IP
  Stud. Adv. Math.}\ }\textbf {\bibinfo {volume} {50}},\ \bibinfo {pages} {347}
  (\bibinfo {year} {2011})},\ \Eprint {http://arxiv.org/abs/1001.2933}
  {arXiv:1001.2933 [hep-th]} \BibitemShut {NoStop}%
\bibitem [{\citenamefont {Andreassen}\ \emph {et~al.}(2016)\citenamefont
  {Andreassen}, \citenamefont {Farhi}, \citenamefont {Frost},\ and\
  \citenamefont {Schwartz}}]{Andreassen:2016cff}%
  \BibitemOpen
  \bibfield  {author} {\bibinfo {author} {\bibfnamefont {A.}~\bibnamefont
  {Andreassen}}, \bibinfo {author} {\bibfnamefont {D.}~\bibnamefont {Farhi}},
  \bibinfo {author} {\bibfnamefont {W.}~\bibnamefont {Frost}}, \ and\ \bibinfo
  {author} {\bibfnamefont {M.~D.}\ \bibnamefont {Schwartz}},\ }\href {\doibase
  10.1103/PhysRevLett.117.231601} {\bibfield  {journal} {\bibinfo  {journal}
  {Phys. Rev. Lett.}\ }\textbf {\bibinfo {volume} {117}},\ \bibinfo {pages}
  {231601} (\bibinfo {year} {2016})},\ \Eprint
  {http://arxiv.org/abs/1602.01102} {arXiv:1602.01102 [hep-th]} \BibitemShut
  {NoStop}%
\bibitem [{\citenamefont {Andreassen}\ \emph {et~al.}(2017)\citenamefont
  {Andreassen}, \citenamefont {Farhi}, \citenamefont {Frost},\ and\
  \citenamefont {Schwartz}}]{Andreassen:2016cvx}%
  \BibitemOpen
  \bibfield  {author} {\bibinfo {author} {\bibfnamefont {A.}~\bibnamefont
  {Andreassen}}, \bibinfo {author} {\bibfnamefont {D.}~\bibnamefont {Farhi}},
  \bibinfo {author} {\bibfnamefont {W.}~\bibnamefont {Frost}}, \ and\ \bibinfo
  {author} {\bibfnamefont {M.~D.}\ \bibnamefont {Schwartz}},\ }\href {\doibase
  10.1103/PhysRevD.95.085011} {\bibfield  {journal} {\bibinfo  {journal} {Phys.
  Rev. D}\ }\textbf {\bibinfo {volume} {95}},\ \bibinfo {pages} {085011}
  (\bibinfo {year} {2017})},\ \Eprint {http://arxiv.org/abs/1604.06090}
  {arXiv:1604.06090 [hep-th]} \BibitemShut {NoStop}%
\bibitem [{\citenamefont {Iliopoulos}\ \emph {et~al.}(1975)\citenamefont
  {Iliopoulos}, \citenamefont {Itzykson},\ and\ \citenamefont
  {Martin}}]{Iliopoulos:1974ur}%
  \BibitemOpen
  \bibfield  {author} {\bibinfo {author} {\bibfnamefont {J.}~\bibnamefont
  {Iliopoulos}}, \bibinfo {author} {\bibfnamefont {C.}~\bibnamefont
  {Itzykson}}, \ and\ \bibinfo {author} {\bibfnamefont {A.}~\bibnamefont
  {Martin}},\ }\href {\doibase 10.1103/RevModPhys.47.165} {\bibfield  {journal}
  {\bibinfo  {journal} {Rev. Mod. Phys.}\ }\textbf {\bibinfo {volume} {47}},\
  \bibinfo {pages} {165} (\bibinfo {year} {1975})}\BibitemShut {NoStop}%
\bibitem [{\citenamefont {Hindmarsh}\ and\ \citenamefont
  {Johnston}(1986)}]{Hindmarsh:1985nc}%
  \BibitemOpen
  \bibfield  {author} {\bibinfo {author} {\bibfnamefont {M.}~\bibnamefont
  {Hindmarsh}}\ and\ \bibinfo {author} {\bibfnamefont {D.}~\bibnamefont
  {Johnston}},\ }\href {\doibase 10.1088/0305-4470/19/1/016} {\bibfield
  {journal} {\bibinfo  {journal} {J. Phys. A}\ }\textbf {\bibinfo {volume}
  {19}},\ \bibinfo {pages} {141} (\bibinfo {year} {1986})}\BibitemShut
  {NoStop}%
\bibitem [{\citenamefont {Fujimoto}\ \emph {et~al.}(1983)\citenamefont
  {Fujimoto}, \citenamefont {O'Raifeartaigh},\ and\ \citenamefont
  {Parravicini}}]{Fujimoto:1982tc}%
  \BibitemOpen
  \bibfield  {author} {\bibinfo {author} {\bibfnamefont {Y.}~\bibnamefont
  {Fujimoto}}, \bibinfo {author} {\bibfnamefont {L.}~\bibnamefont
  {O'Raifeartaigh}}, \ and\ \bibinfo {author} {\bibfnamefont {G.}~\bibnamefont
  {Parravicini}},\ }\href {\doibase 10.1016/0550-3213(83)90305-X} {\bibfield
  {journal} {\bibinfo  {journal} {Nucl. Phys. B}\ }\textbf {\bibinfo {volume}
  {212}},\ \bibinfo {pages} {268} (\bibinfo {year} {1983})}\BibitemShut
  {NoStop}%
\bibitem [{\citenamefont {Fukuda}\ and\ \citenamefont
  {Kyriakopoulos}(1975)}]{Fukuda:1974ey}%
  \BibitemOpen
  \bibfield  {author} {\bibinfo {author} {\bibfnamefont {R.}~\bibnamefont
  {Fukuda}}\ and\ \bibinfo {author} {\bibfnamefont {E.}~\bibnamefont
  {Kyriakopoulos}},\ }\href {\doibase 10.1016/0550-3213(75)90014-0} {\bibfield
  {journal} {\bibinfo  {journal} {Nucl. Phys. B}\ }\textbf {\bibinfo {volume}
  {85}},\ \bibinfo {pages} {354} (\bibinfo {year} {1975})}\BibitemShut
  {NoStop}%
\bibitem [{\citenamefont {Weinberg}\ and\ \citenamefont
  {Wu}(1987)}]{Weinberg:1987vp}%
  \BibitemOpen
  \bibfield  {author} {\bibinfo {author} {\bibfnamefont {E.~J.}\ \bibnamefont
  {Weinberg}}\ and\ \bibinfo {author} {\bibfnamefont {A.-q.}\ \bibnamefont
  {Wu}},\ }\href {\doibase 10.1103/PhysRevD.36.2474} {\bibfield  {journal}
  {\bibinfo  {journal} {Phys. Rev. D}\ }\textbf {\bibinfo {volume} {36}},\
  \bibinfo {pages} {2474} (\bibinfo {year} {1987})}\BibitemShut {NoStop}%
\bibitem [{\citenamefont {Plascencia}\ and\ \citenamefont
  {Tamarit}(2016)}]{Plascencia:2015pga}%
  \BibitemOpen
  \bibfield  {author} {\bibinfo {author} {\bibfnamefont {A.~D.}\ \bibnamefont
  {Plascencia}}\ and\ \bibinfo {author} {\bibfnamefont {C.}~\bibnamefont
  {Tamarit}},\ }\href {\doibase 10.1007/JHEP10(2016)099} {\bibfield  {journal}
  {\bibinfo  {journal} {JHEP}\ }\textbf {\bibinfo {volume} {10}},\ \bibinfo
  {pages} {099} (\bibinfo {year} {2016})},\ \Eprint
  {http://arxiv.org/abs/1510.07613} {arXiv:1510.07613 [hep-ph]} \BibitemShut
  {NoStop}%
\bibitem [{\citenamefont {Ginsparg}(1980)}]{Ginsparg:1980ef}%
  \BibitemOpen
  \bibfield  {author} {\bibinfo {author} {\bibfnamefont {P.~H.}\ \bibnamefont
  {Ginsparg}},\ }\href {\doibase 10.1016/0550-3213(80)90418-6} {\bibfield
  {journal} {\bibinfo  {journal} {Nucl. Phys. B}\ }\textbf {\bibinfo {volume}
  {170}},\ \bibinfo {pages} {388} (\bibinfo {year} {1980})}\BibitemShut
  {NoStop}%
\bibitem [{\citenamefont {Appelquist}\ and\ \citenamefont
  {Pisarski}(1981)}]{Appelquist:1981vg}%
  \BibitemOpen
  \bibfield  {author} {\bibinfo {author} {\bibfnamefont {T.}~\bibnamefont
  {Appelquist}}\ and\ \bibinfo {author} {\bibfnamefont {R.~D.}\ \bibnamefont
  {Pisarski}},\ }\href {\doibase 10.1103/PhysRevD.23.2305} {\bibfield
  {journal} {\bibinfo  {journal} {Phys. Rev. D}\ }\textbf {\bibinfo {volume}
  {23}},\ \bibinfo {pages} {2305} (\bibinfo {year} {1981})}\BibitemShut
  {NoStop}%
\bibitem [{\citenamefont {Nadkarni}(1983)}]{Nadkarni:1982kb}%
  \BibitemOpen
  \bibfield  {author} {\bibinfo {author} {\bibfnamefont {S.}~\bibnamefont
  {Nadkarni}},\ }\href {\doibase 10.1103/PhysRevD.27.917} {\bibfield  {journal}
  {\bibinfo  {journal} {Phys. Rev. D}\ }\textbf {\bibinfo {volume} {27}},\
  \bibinfo {pages} {917} (\bibinfo {year} {1983})}\BibitemShut {NoStop}%
\bibitem [{\citenamefont {Farakos}\ \emph {et~al.}(1994)\citenamefont
  {Farakos}, \citenamefont {Kajantie}, \citenamefont {Rummukainen},\ and\
  \citenamefont {Shaposhnikov}}]{Farakos:1994kx}%
  \BibitemOpen
  \bibfield  {author} {\bibinfo {author} {\bibfnamefont {K.}~\bibnamefont
  {Farakos}}, \bibinfo {author} {\bibfnamefont {K.}~\bibnamefont {Kajantie}},
  \bibinfo {author} {\bibfnamefont {K.}~\bibnamefont {Rummukainen}}, \ and\
  \bibinfo {author} {\bibfnamefont {M.~E.}\ \bibnamefont {Shaposhnikov}},\
  }\href {\doibase 10.1016/0550-3213(94)90173-2} {\bibfield  {journal}
  {\bibinfo  {journal} {Nucl. Phys. B}\ }\textbf {\bibinfo {volume} {425}},\
  \bibinfo {pages} {67} (\bibinfo {year} {1994})},\ \Eprint
  {http://arxiv.org/abs/hep-ph/9404201} {arXiv:hep-ph/9404201} \BibitemShut
  {NoStop}%
\bibitem [{\citenamefont {Braaten}\ and\ \citenamefont
  {Nieto}(1995)}]{Braaten:1995cm}%
  \BibitemOpen
  \bibfield  {author} {\bibinfo {author} {\bibfnamefont {E.}~\bibnamefont
  {Braaten}}\ and\ \bibinfo {author} {\bibfnamefont {A.}~\bibnamefont
  {Nieto}},\ }\href {\doibase 10.1103/PhysRevD.51.6990} {\bibfield  {journal}
  {\bibinfo  {journal} {Phys. Rev. D}\ }\textbf {\bibinfo {volume} {51}},\
  \bibinfo {pages} {6990} (\bibinfo {year} {1995})},\ \Eprint
  {http://arxiv.org/abs/hep-ph/9501375} {arXiv:hep-ph/9501375} \BibitemShut
  {NoStop}%
\bibitem [{\citenamefont {Kajantie}\ \emph {et~al.}(1996)\citenamefont
  {Kajantie}, \citenamefont {Laine}, \citenamefont {Rummukainen},\ and\
  \citenamefont {Shaposhnikov}}]{Kajantie:1995dw}%
  \BibitemOpen
  \bibfield  {author} {\bibinfo {author} {\bibfnamefont {K.}~\bibnamefont
  {Kajantie}}, \bibinfo {author} {\bibfnamefont {M.}~\bibnamefont {Laine}},
  \bibinfo {author} {\bibfnamefont {K.}~\bibnamefont {Rummukainen}}, \ and\
  \bibinfo {author} {\bibfnamefont {M.~E.}\ \bibnamefont {Shaposhnikov}},\
  }\href {\doibase 10.1016/0550-3213(95)00549-8} {\bibfield  {journal}
  {\bibinfo  {journal} {Nucl. Phys. B}\ }\textbf {\bibinfo {volume} {458}},\
  \bibinfo {pages} {90} (\bibinfo {year} {1996})},\ \Eprint
  {http://arxiv.org/abs/hep-ph/9508379} {arXiv:hep-ph/9508379} \BibitemShut
  {NoStop}%
\bibitem [{\citenamefont {Croon}\ \emph {et~al.}(2021)\citenamefont {Croon},
  \citenamefont {Gould}, \citenamefont {Schicho}, \citenamefont {Tenkanen},\
  and\ \citenamefont {White}}]{Croon:2020cgk}%
  \BibitemOpen
  \bibfield  {author} {\bibinfo {author} {\bibfnamefont {D.}~\bibnamefont
  {Croon}}, \bibinfo {author} {\bibfnamefont {O.}~\bibnamefont {Gould}},
  \bibinfo {author} {\bibfnamefont {P.}~\bibnamefont {Schicho}}, \bibinfo
  {author} {\bibfnamefont {T.~V.~I.}\ \bibnamefont {Tenkanen}}, \ and\ \bibinfo
  {author} {\bibfnamefont {G.}~\bibnamefont {White}},\ }\href {\doibase
  10.1007/JHEP04(2021)055} {\bibfield  {journal} {\bibinfo  {journal} {JHEP}\
  }\textbf {\bibinfo {volume} {04}},\ \bibinfo {pages} {055} (\bibinfo {year}
  {2021})},\ \Eprint {http://arxiv.org/abs/2009.10080} {arXiv:2009.10080
  [hep-ph]} \BibitemShut {NoStop}%
\bibitem [{\citenamefont {Alford}\ and\ \citenamefont
  {March-Russell}(1994)}]{Alford:1993br}%
  \BibitemOpen
  \bibfield  {author} {\bibinfo {author} {\bibfnamefont {M.~G.}\ \bibnamefont
  {Alford}}\ and\ \bibinfo {author} {\bibfnamefont {J.}~\bibnamefont
  {March-Russell}},\ }\href {\doibase 10.1016/0550-3213(94)90483-9} {\bibfield
  {journal} {\bibinfo  {journal} {Nucl. Phys. B}\ }\textbf {\bibinfo {volume}
  {417}},\ \bibinfo {pages} {527} (\bibinfo {year} {1994})},\ \Eprint
  {http://arxiv.org/abs/hep-ph/9308364} {arXiv:hep-ph/9308364} \BibitemShut
  {NoStop}%
\bibitem [{\citenamefont {Litim}(1997)}]{Litim:1996nw}%
  \BibitemOpen
  \bibfield  {author} {\bibinfo {author} {\bibfnamefont {D.~F.}\ \bibnamefont
  {Litim}},\ }\href {\doibase 10.1016/S0370-2693(96)01613-9} {\bibfield
  {journal} {\bibinfo  {journal} {Phys. Lett. B}\ }\textbf {\bibinfo {volume}
  {393}},\ \bibinfo {pages} {103} (\bibinfo {year} {1997})},\ \Eprint
  {http://arxiv.org/abs/hep-th/9609040} {arXiv:hep-th/9609040} \BibitemShut
  {NoStop}%
\bibitem [{\citenamefont {Berges}\ \emph {et~al.}(1997)\citenamefont {Berges},
  \citenamefont {Tetradis},\ and\ \citenamefont {Wetterich}}]{Berges:1996ib}%
  \BibitemOpen
  \bibfield  {author} {\bibinfo {author} {\bibfnamefont {J.}~\bibnamefont
  {Berges}}, \bibinfo {author} {\bibfnamefont {N.}~\bibnamefont {Tetradis}}, \
  and\ \bibinfo {author} {\bibfnamefont {C.}~\bibnamefont {Wetterich}},\ }\href
  {\doibase 10.1016/S0370-2693(96)01654-1} {\bibfield  {journal} {\bibinfo
  {journal} {Phys. Lett. B}\ }\textbf {\bibinfo {volume} {393}},\ \bibinfo
  {pages} {387} (\bibinfo {year} {1997})},\ \Eprint
  {http://arxiv.org/abs/hep-ph/9610354} {arXiv:hep-ph/9610354} \BibitemShut
  {NoStop}%
\bibitem [{\citenamefont {Freire}\ and\ \citenamefont
  {Litim}(2001)}]{Freire:2000sx}%
  \BibitemOpen
  \bibfield  {author} {\bibinfo {author} {\bibfnamefont {F.}~\bibnamefont
  {Freire}}\ and\ \bibinfo {author} {\bibfnamefont {D.~F.}\ \bibnamefont
  {Litim}},\ }\href {\doibase 10.1103/PhysRevD.64.045014} {\bibfield  {journal}
  {\bibinfo  {journal} {Phys. Rev. D}\ }\textbf {\bibinfo {volume} {64}},\
  \bibinfo {pages} {045014} (\bibinfo {year} {2001})},\ \Eprint
  {http://arxiv.org/abs/hep-ph/0002153} {arXiv:hep-ph/0002153} \BibitemShut
  {NoStop}%
\bibitem [{\citenamefont {Berges}\ \emph {et~al.}(2002)\citenamefont {Berges},
  \citenamefont {Tetradis},\ and\ \citenamefont {Wetterich}}]{Berges:2000ew}%
  \BibitemOpen
  \bibfield  {author} {\bibinfo {author} {\bibfnamefont {J.}~\bibnamefont
  {Berges}}, \bibinfo {author} {\bibfnamefont {N.}~\bibnamefont {Tetradis}}, \
  and\ \bibinfo {author} {\bibfnamefont {C.}~\bibnamefont {Wetterich}},\ }\href
  {\doibase 10.1016/S0370-1573(01)00098-9} {\bibfield  {journal} {\bibinfo
  {journal} {Phys. Rept.}\ }\textbf {\bibinfo {volume} {363}},\ \bibinfo
  {pages} {223} (\bibinfo {year} {2002})},\ \Eprint
  {http://arxiv.org/abs/hep-ph/0005122} {arXiv:hep-ph/0005122} \BibitemShut
  {NoStop}%
\bibitem [{\citenamefont {Eichhorn}\ \emph {et~al.}(2021)\citenamefont
  {Eichhorn}, \citenamefont {Lumma}, \citenamefont {Pawlowski}, \citenamefont
  {Reichert},\ and\ \citenamefont {Yamada}}]{Eichhorn:2020upj}%
  \BibitemOpen
  \bibfield  {author} {\bibinfo {author} {\bibfnamefont {A.}~\bibnamefont
  {Eichhorn}}, \bibinfo {author} {\bibfnamefont {J.}~\bibnamefont {Lumma}},
  \bibinfo {author} {\bibfnamefont {J.~M.}\ \bibnamefont {Pawlowski}}, \bibinfo
  {author} {\bibfnamefont {M.}~\bibnamefont {Reichert}}, \ and\ \bibinfo
  {author} {\bibfnamefont {M.}~\bibnamefont {Yamada}},\ }\href {\doibase
  10.1088/1475-7516/2021/05/006} {\bibfield  {journal} {\bibinfo  {journal}
  {JCAP}\ }\textbf {\bibinfo {volume} {05}},\ \bibinfo {pages} {006} (\bibinfo
  {year} {2021})},\ \Eprint {http://arxiv.org/abs/2010.00017} {arXiv:2010.00017
  [hep-ph]} \BibitemShut {NoStop}%
\bibitem [{\citenamefont {Andreassen}\ \emph {et~al.}(2018)\citenamefont
  {Andreassen}, \citenamefont {Frost},\ and\ \citenamefont
  {Schwartz}}]{Andreassen:2017rzq}%
  \BibitemOpen
  \bibfield  {author} {\bibinfo {author} {\bibfnamefont {A.}~\bibnamefont
  {Andreassen}}, \bibinfo {author} {\bibfnamefont {W.}~\bibnamefont {Frost}}, \
  and\ \bibinfo {author} {\bibfnamefont {M.~D.}\ \bibnamefont {Schwartz}},\
  }\href {\doibase 10.1103/PhysRevD.97.056006} {\bibfield  {journal} {\bibinfo
  {journal} {Phys. Rev. D}\ }\textbf {\bibinfo {volume} {97}},\ \bibinfo
  {pages} {056006} (\bibinfo {year} {2018})},\ \Eprint
  {http://arxiv.org/abs/1707.08124} {arXiv:1707.08124 [hep-ph]} \BibitemShut
  {NoStop}%
\bibitem [{\citenamefont {Croon}\ \emph
  {et~al.}(2023{\natexlab{a}})\citenamefont {Croon}, \citenamefont {Hall},\
  and\ \citenamefont {Schwartz}}]{direct-forthcoming}%
  \BibitemOpen
  \bibfield  {author} {\bibinfo {author} {\bibfnamefont {D.}~\bibnamefont
  {Croon}}, \bibinfo {author} {\bibfnamefont {E.}~\bibnamefont {Hall}}, \ and\
  \bibinfo {author} {\bibfnamefont {M.}~\bibnamefont {Schwartz}},\ }\href@noop
  {} {\bibfield  {journal} {\bibinfo  {journal} {forthcoming work}\ } (\bibinfo
  {year} {2023}{\natexlab{a}})}\BibitemShut {NoStop}%
\bibitem [{\citenamefont {Litim}\ \emph {et~al.}(2006)\citenamefont {Litim},
  \citenamefont {Pawlowski},\ and\ \citenamefont {Vergara}}]{Litim:2006nn}%
  \BibitemOpen
  \bibfield  {author} {\bibinfo {author} {\bibfnamefont {D.~F.}\ \bibnamefont
  {Litim}}, \bibinfo {author} {\bibfnamefont {J.~M.}\ \bibnamefont
  {Pawlowski}}, \ and\ \bibinfo {author} {\bibfnamefont {L.}~\bibnamefont
  {Vergara}},\ }\href@noop {} {\  (\bibinfo {year} {2006})},\ \Eprint
  {http://arxiv.org/abs/hep-th/0602140} {arXiv:hep-th/0602140} \BibitemShut
  {NoStop}%
\bibitem [{\citenamefont {Litim}(2002)}]{Litim:2002cf}%
  \BibitemOpen
  \bibfield  {author} {\bibinfo {author} {\bibfnamefont {D.~F.}\ \bibnamefont
  {Litim}},\ }\href {\doibase 10.1016/S0550-3213(02)00186-4} {\bibfield
  {journal} {\bibinfo  {journal} {Nucl. Phys. B}\ }\textbf {\bibinfo {volume}
  {631}},\ \bibinfo {pages} {128} (\bibinfo {year} {2002})},\ \Eprint
  {http://arxiv.org/abs/hep-th/0203006} {arXiv:hep-th/0203006} \BibitemShut
  {NoStop}%
\bibitem [{\citenamefont {Litim}(2000)}]{Litim:2000ci}%
  \BibitemOpen
  \bibfield  {author} {\bibinfo {author} {\bibfnamefont {D.~F.}\ \bibnamefont
  {Litim}},\ }\href {\doibase 10.1016/S0370-2693(00)00748-6} {\bibfield
  {journal} {\bibinfo  {journal} {Phys. Lett. B}\ }\textbf {\bibinfo {volume}
  {486}},\ \bibinfo {pages} {92} (\bibinfo {year} {2000})},\ \Eprint
  {http://arxiv.org/abs/hep-th/0005245} {arXiv:hep-th/0005245} \BibitemShut
  {NoStop}%
\bibitem [{\citenamefont {Litim}(2001)}]{Litim:2001up}%
  \BibitemOpen
  \bibfield  {author} {\bibinfo {author} {\bibfnamefont {D.~F.}\ \bibnamefont
  {Litim}},\ }\href {\doibase 10.1103/PhysRevD.64.105007} {\bibfield  {journal}
  {\bibinfo  {journal} {Phys. Rev. D}\ }\textbf {\bibinfo {volume} {64}},\
  \bibinfo {pages} {105007} (\bibinfo {year} {2001})},\ \Eprint
  {http://arxiv.org/abs/hep-th/0103195} {arXiv:hep-th/0103195} \BibitemShut
  {NoStop}%
\bibitem [{\citenamefont {Garbrecht}\ and\ \citenamefont
  {Millington}(2016)}]{Garbrecht:2015cla}%
  \BibitemOpen
  \bibfield  {author} {\bibinfo {author} {\bibfnamefont {B.}~\bibnamefont
  {Garbrecht}}\ and\ \bibinfo {author} {\bibfnamefont {P.}~\bibnamefont
  {Millington}},\ }\href {\doibase 10.1016/j.nuclphysb.2016.02.022} {\bibfield
  {journal} {\bibinfo  {journal} {Nucl. Phys. B}\ }\textbf {\bibinfo {volume}
  {906}},\ \bibinfo {pages} {105} (\bibinfo {year} {2016})},\ \Eprint
  {http://arxiv.org/abs/1509.07847} {arXiv:1509.07847 [hep-th]} \BibitemShut
  {NoStop}%
\bibitem [{\citenamefont {Garbrecht}\ and\ \citenamefont
  {Millington}(2015)}]{Garbrecht:2015yza}%
  \BibitemOpen
  \bibfield  {author} {\bibinfo {author} {\bibfnamefont {B.}~\bibnamefont
  {Garbrecht}}\ and\ \bibinfo {author} {\bibfnamefont {P.}~\bibnamefont
  {Millington}},\ }\href {\doibase 10.1103/PhysRevD.92.125022} {\bibfield
  {journal} {\bibinfo  {journal} {Phys. Rev. D}\ }\textbf {\bibinfo {volume}
  {92}},\ \bibinfo {pages} {125022} (\bibinfo {year} {2015})},\ \Eprint
  {http://arxiv.org/abs/1509.08480} {arXiv:1509.08480 [hep-ph]} \BibitemShut
  {NoStop}%
\bibitem [{\citenamefont {Alexander}\ \emph
  {et~al.}(2019{\natexlab{a}})\citenamefont {Alexander}, \citenamefont
  {Millington}, \citenamefont {Nursey},\ and\ \citenamefont
  {Saffin}}]{Alexander:2019cgw}%
  \BibitemOpen
  \bibfield  {author} {\bibinfo {author} {\bibfnamefont {E.}~\bibnamefont
  {Alexander}}, \bibinfo {author} {\bibfnamefont {P.}~\bibnamefont
  {Millington}}, \bibinfo {author} {\bibfnamefont {J.}~\bibnamefont {Nursey}},
  \ and\ \bibinfo {author} {\bibfnamefont {P.~M.}\ \bibnamefont {Saffin}},\
  }\href {\doibase 10.1103/PhysRevD.100.101702} {\bibfield  {journal} {\bibinfo
   {journal} {Phys. Rev. D}\ }\textbf {\bibinfo {volume} {100}},\ \bibinfo
  {pages} {101702(R)} (\bibinfo {year} {2019}{\natexlab{a}})},\ \Eprint
  {http://arxiv.org/abs/1907.06503} {arXiv:1907.06503 [hep-th]} \BibitemShut
  {NoStop}%
\bibitem [{\citenamefont {Alexander}\ \emph
  {et~al.}(2019{\natexlab{b}})\citenamefont {Alexander}, \citenamefont
  {Millington}, \citenamefont {Nursey},\ and\ \citenamefont
  {Saffin}}]{Alexander:2019quf}%
  \BibitemOpen
  \bibfield  {author} {\bibinfo {author} {\bibfnamefont {E.}~\bibnamefont
  {Alexander}}, \bibinfo {author} {\bibfnamefont {P.}~\bibnamefont
  {Millington}}, \bibinfo {author} {\bibfnamefont {J.}~\bibnamefont {Nursey}},
  \ and\ \bibinfo {author} {\bibfnamefont {P.~M.}\ \bibnamefont {Saffin}},\
  }\href@noop {} {\  (\bibinfo {year} {2019}{\natexlab{b}})},\ \Eprint
  {http://arxiv.org/abs/1908.02214} {arXiv:1908.02214 [hep-th]} \BibitemShut
  {NoStop}%
\bibitem [{\citenamefont {Croon}\ \emph
  {et~al.}(2023{\natexlab{b}})\citenamefont {Croon}, \citenamefont {Hall},\
  and\ \citenamefont {Millington}}]{2PI-forthcoming}%
  \BibitemOpen
  \bibfield  {author} {\bibinfo {author} {\bibfnamefont {D.}~\bibnamefont
  {Croon}}, \bibinfo {author} {\bibfnamefont {E.}~\bibnamefont {Hall}}, \ and\
  \bibinfo {author} {\bibfnamefont {P.}~\bibnamefont {Millington}},\
  }\href@noop {} {\bibfield  {journal} {\bibinfo  {journal} {forthcoming work}\
  } (\bibinfo {year} {2023}{\natexlab{b}})}\BibitemShut {NoStop}%
\bibitem [{\citenamefont {Gies}\ and\ \citenamefont
  {Sondenheimer}(2018)}]{Gies:2017ajd}%
  \BibitemOpen
  \bibfield  {author} {\bibinfo {author} {\bibfnamefont {H.}~\bibnamefont
  {Gies}}\ and\ \bibinfo {author} {\bibfnamefont {R.}~\bibnamefont
  {Sondenheimer}},\ }\href {\doibase 10.1098/rsta.2017.0120} {\bibfield
  {journal} {\bibinfo  {journal} {Phil. Trans. Roy. Soc. Lond. A}\ }\textbf
  {\bibinfo {volume} {376}},\ \bibinfo {pages} {20170120} (\bibinfo {year}
  {2018})},\ \Eprint {http://arxiv.org/abs/1708.04305} {arXiv:1708.04305
  [hep-ph]} \BibitemShut {NoStop}%
\bibitem [{\citenamefont {Fukushima}(2004)}]{Fukushima:2003fw}%
  \BibitemOpen
  \bibfield  {author} {\bibinfo {author} {\bibfnamefont {K.}~\bibnamefont
  {Fukushima}},\ }\href {\doibase 10.1016/j.physletb.2004.04.027} {\bibfield
  {journal} {\bibinfo  {journal} {Phys. Lett. B}\ }\textbf {\bibinfo {volume}
  {591}},\ \bibinfo {pages} {277} (\bibinfo {year} {2004})},\ \Eprint
  {http://arxiv.org/abs/hep-ph/0310121} {arXiv:hep-ph/0310121} \BibitemShut
  {NoStop}%
\bibitem [{\citenamefont {Schaefer}\ and\ \citenamefont
  {Pirner}(1999)}]{Schaefer:1999em}%
  \BibitemOpen
  \bibfield  {author} {\bibinfo {author} {\bibfnamefont {B.-J.}\ \bibnamefont
  {Schaefer}}\ and\ \bibinfo {author} {\bibfnamefont {H.-J.}\ \bibnamefont
  {Pirner}},\ }\href {\doibase 10.1016/S0375-9474(99)00409-1} {\bibfield
  {journal} {\bibinfo  {journal} {Nucl. Phys. A}\ }\textbf {\bibinfo {volume}
  {660}},\ \bibinfo {pages} {439} (\bibinfo {year} {1999})},\ \Eprint
  {http://arxiv.org/abs/nucl-th/9903003} {arXiv:nucl-th/9903003} \BibitemShut
  {NoStop}%
\bibitem [{\citenamefont {Bohr}\ \emph {et~al.}(2001)\citenamefont {Bohr},
  \citenamefont {Schaefer},\ and\ \citenamefont {Wambach}}]{Bohr:2000gp}%
  \BibitemOpen
  \bibfield  {author} {\bibinfo {author} {\bibfnamefont {O.}~\bibnamefont
  {Bohr}}, \bibinfo {author} {\bibfnamefont {B.~J.}\ \bibnamefont {Schaefer}},
  \ and\ \bibinfo {author} {\bibfnamefont {J.}~\bibnamefont {Wambach}},\ }\href
  {\doibase 10.1142/S0217751X0100502X} {\bibfield  {journal} {\bibinfo
  {journal} {Int. J. Mod. Phys. A}\ }\textbf {\bibinfo {volume} {16}},\
  \bibinfo {pages} {3823} (\bibinfo {year} {2001})},\ \Eprint
  {http://arxiv.org/abs/hep-ph/0007098} {arXiv:hep-ph/0007098} \BibitemShut
  {NoStop}%
\bibitem [{\citenamefont {Hochberg}\ \emph {et~al.}(2015)\citenamefont
  {Hochberg}, \citenamefont {Kuflik}, \citenamefont {Murayama}, \citenamefont
  {Volansky},\ and\ \citenamefont {Wacker}}]{Hochberg:2014kqa}%
  \BibitemOpen
  \bibfield  {author} {\bibinfo {author} {\bibfnamefont {Y.}~\bibnamefont
  {Hochberg}}, \bibinfo {author} {\bibfnamefont {E.}~\bibnamefont {Kuflik}},
  \bibinfo {author} {\bibfnamefont {H.}~\bibnamefont {Murayama}}, \bibinfo
  {author} {\bibfnamefont {T.}~\bibnamefont {Volansky}}, \ and\ \bibinfo
  {author} {\bibfnamefont {J.~G.}\ \bibnamefont {Wacker}},\ }\href {\doibase
  10.1103/PhysRevLett.115.021301} {\bibfield  {journal} {\bibinfo  {journal}
  {Phys. Rev. Lett.}\ }\textbf {\bibinfo {volume} {115}},\ \bibinfo {pages}
  {021301} (\bibinfo {year} {2015})},\ \Eprint {http://arxiv.org/abs/1411.3727}
  {arXiv:1411.3727 [hep-ph]} \BibitemShut {NoStop}%
\bibitem [{\citenamefont {Tsumura}\ \emph {et~al.}(2017)\citenamefont
  {Tsumura}, \citenamefont {Yamada},\ and\ \citenamefont
  {Yamaguchi}}]{Tsumura:2017knk}%
  \BibitemOpen
  \bibfield  {author} {\bibinfo {author} {\bibfnamefont {K.}~\bibnamefont
  {Tsumura}}, \bibinfo {author} {\bibfnamefont {M.}~\bibnamefont {Yamada}}, \
  and\ \bibinfo {author} {\bibfnamefont {Y.}~\bibnamefont {Yamaguchi}},\ }\href
  {\doibase 10.1088/1475-7516/2017/07/044} {\bibfield  {journal} {\bibinfo
  {journal} {JCAP}\ }\textbf {\bibinfo {volume} {07}},\ \bibinfo {pages} {044}
  (\bibinfo {year} {2017})},\ \Eprint {http://arxiv.org/abs/1704.00219}
  {arXiv:1704.00219 [hep-ph]} \BibitemShut {NoStop}%
\bibitem [{\citenamefont {Aoki}\ \emph {et~al.}(2017)\citenamefont {Aoki},
  \citenamefont {Goto},\ and\ \citenamefont {Kubo}}]{Aoki:2017aws}%
  \BibitemOpen
  \bibfield  {author} {\bibinfo {author} {\bibfnamefont {M.}~\bibnamefont
  {Aoki}}, \bibinfo {author} {\bibfnamefont {H.}~\bibnamefont {Goto}}, \ and\
  \bibinfo {author} {\bibfnamefont {J.}~\bibnamefont {Kubo}},\ }\href {\doibase
  10.1103/PhysRevD.96.075045} {\bibfield  {journal} {\bibinfo  {journal} {Phys.
  Rev. D}\ }\textbf {\bibinfo {volume} {96}},\ \bibinfo {pages} {075045}
  (\bibinfo {year} {2017})},\ \Eprint {http://arxiv.org/abs/1709.07572}
  {arXiv:1709.07572 [hep-ph]} \BibitemShut {NoStop}%
\bibitem [{\citenamefont {Bai}\ \emph {et~al.}(2019)\citenamefont {Bai},
  \citenamefont {Long},\ and\ \citenamefont {Lu}}]{Bai:2018dxf}%
  \BibitemOpen
  \bibfield  {author} {\bibinfo {author} {\bibfnamefont {Y.}~\bibnamefont
  {Bai}}, \bibinfo {author} {\bibfnamefont {A.~J.}\ \bibnamefont {Long}}, \
  and\ \bibinfo {author} {\bibfnamefont {S.}~\bibnamefont {Lu}},\ }\href
  {\doibase 10.1103/PhysRevD.99.055047} {\bibfield  {journal} {\bibinfo
  {journal} {Phys. Rev. D}\ }\textbf {\bibinfo {volume} {99}},\ \bibinfo
  {pages} {055047} (\bibinfo {year} {2019})},\ \Eprint
  {http://arxiv.org/abs/1810.04360} {arXiv:1810.04360 [hep-ph]} \BibitemShut
  {NoStop}%
\bibitem [{\citenamefont {Helmboldt}\ \emph {et~al.}(2019)\citenamefont
  {Helmboldt}, \citenamefont {Kubo},\ and\ \citenamefont {van~der
  Woude}}]{Helmboldt:2019pan}%
  \BibitemOpen
  \bibfield  {author} {\bibinfo {author} {\bibfnamefont {A.~J.}\ \bibnamefont
  {Helmboldt}}, \bibinfo {author} {\bibfnamefont {J.}~\bibnamefont {Kubo}}, \
  and\ \bibinfo {author} {\bibfnamefont {S.}~\bibnamefont {van~der Woude}},\
  }\href {\doibase 10.1103/PhysRevD.100.055025} {\bibfield  {journal} {\bibinfo
   {journal} {Phys. Rev. D}\ }\textbf {\bibinfo {volume} {100}},\ \bibinfo
  {pages} {055025} (\bibinfo {year} {2019})},\ \Eprint
  {http://arxiv.org/abs/1904.07891} {arXiv:1904.07891 [hep-ph]} \BibitemShut
  {NoStop}%
\bibitem [{\citenamefont {Hall}\ \emph {et~al.}(2023)\citenamefont {Hall},
  \citenamefont {Konstandin}, \citenamefont {McGehee},\ and\ \citenamefont
  {Murayama}}]{Hall:2019rld}%
  \BibitemOpen
  \bibfield  {author} {\bibinfo {author} {\bibfnamefont {E.}~\bibnamefont
  {Hall}}, \bibinfo {author} {\bibfnamefont {T.}~\bibnamefont {Konstandin}},
  \bibinfo {author} {\bibfnamefont {R.}~\bibnamefont {McGehee}}, \ and\
  \bibinfo {author} {\bibfnamefont {H.}~\bibnamefont {Murayama}},\ }\href
  {\doibase 10.1103/PhysRevD.107.055011} {\bibfield  {journal} {\bibinfo
  {journal} {Phys. Rev. D}\ }\textbf {\bibinfo {volume} {107}},\ \bibinfo
  {pages} {055011} (\bibinfo {year} {2023})},\ \Eprint
  {http://arxiv.org/abs/1911.12342} {arXiv:1911.12342 [hep-ph]} \BibitemShut
  {NoStop}%
\bibitem [{\citenamefont {Croon}\ \emph {et~al.}(2019)\citenamefont {Croon},
  \citenamefont {Houtz},\ and\ \citenamefont {Sanz}}]{Croon:2019iuh}%
  \BibitemOpen
  \bibfield  {author} {\bibinfo {author} {\bibfnamefont {D.}~\bibnamefont
  {Croon}}, \bibinfo {author} {\bibfnamefont {R.}~\bibnamefont {Houtz}}, \ and\
  \bibinfo {author} {\bibfnamefont {V.}~\bibnamefont {Sanz}},\ }\href {\doibase
  10.1007/JHEP07(2019)146} {\bibfield  {journal} {\bibinfo  {journal} {JHEP}\
  }\textbf {\bibinfo {volume} {07}},\ \bibinfo {pages} {146} (\bibinfo {year}
  {2019})},\ \Eprint {http://arxiv.org/abs/1904.10967} {arXiv:1904.10967
  [hep-ph]} \BibitemShut {NoStop}%
\bibitem [{\citenamefont {Ipek}\ and\ \citenamefont
  {Tait}(2019)}]{Ipek:2018lhm}%
  \BibitemOpen
  \bibfield  {author} {\bibinfo {author} {\bibfnamefont {S.}~\bibnamefont
  {Ipek}}\ and\ \bibinfo {author} {\bibfnamefont {T.~M.~P.}\ \bibnamefont
  {Tait}},\ }\href {\doibase 10.1103/PhysRevLett.122.112001} {\bibfield
  {journal} {\bibinfo  {journal} {Phys. Rev. Lett.}\ }\textbf {\bibinfo
  {volume} {122}},\ \bibinfo {pages} {112001} (\bibinfo {year} {2019})},\
  \Eprint {http://arxiv.org/abs/1811.00559} {arXiv:1811.00559 [hep-ph]}
  \BibitemShut {NoStop}%
\bibitem [{\citenamefont {Croon}\ \emph {et~al.}(2020)\citenamefont {Croon},
  \citenamefont {Howard}, \citenamefont {Ipek},\ and\ \citenamefont
  {Tait}}]{Croon:2019ugf}%
  \BibitemOpen
  \bibfield  {author} {\bibinfo {author} {\bibfnamefont {D.}~\bibnamefont
  {Croon}}, \bibinfo {author} {\bibfnamefont {J.~N.}\ \bibnamefont {Howard}},
  \bibinfo {author} {\bibfnamefont {S.}~\bibnamefont {Ipek}}, \ and\ \bibinfo
  {author} {\bibfnamefont {T.~M.~P.}\ \bibnamefont {Tait}},\ }\href {\doibase
  10.1103/PhysRevD.101.055042} {\bibfield  {journal} {\bibinfo  {journal}
  {Phys. Rev. D}\ }\textbf {\bibinfo {volume} {101}},\ \bibinfo {pages}
  {055042} (\bibinfo {year} {2020})},\ \Eprint
  {http://arxiv.org/abs/1911.01432} {arXiv:1911.01432 [hep-ph]} \BibitemShut
  {NoStop}%
\end{thebibliography}%
\end{document}